\documentclass[aps,prb,twocolumn,citeautoscript]{revtex4-1}

\bibpunct{[}{]}{,}{n}{}{}

\usepackage{hyperref}

\usepackage{multirow}
\usepackage{graphicx}
\usepackage{longtable}
\usepackage[utf8]{inputenc}
\usepackage{epstopdf}
\usepackage{longtable}
\usepackage{textcomp}
\usepackage[usenames,dvipsnames]{color}
\usepackage{amsbsy}
\usepackage{amsmath}
\usepackage{amsfonts}
\usepackage{mathtools}
\usepackage{braket}
\usepackage{array}
\usepackage{gensymb}
\usepackage[normalem]{ulem}
\usepackage{placeins}

\newcommand{\eg}{e.g.,\@ }
\newcommand{\Eg}{\textit{E.g.},\@ }
\newcommand{\ie}{i.e.\@ }

\newcommand{\vc}[1]{\boldsymbol{\mathrm #1}}
\newcommand{\op}[1]{\mathcal{#1}}
\newcommand{\vop}[1]{\boldsymbol{\mathcal{#1}}}
\newcommand{\commutator}[2]{[\,#1\,,\,#2\,]}
\newcommand\varpm{\mathbin{\vcenter{\hbox{%
  \oalign{\hfil$\scriptstyle+$\hfil\cr
          \noalign{\kern-.3ex}
          $\scriptscriptstyle({-})$\cr}%
}}}}

\makeatletter
  \def\my@tag@font{\normalsize}
  \def\maketag@@@#1{\hbox{\m@th\normalfont\my@tag@font#1}}
  \let\amsmath@eqref\eqref
  \renewcommand\eqref[1]{{\let\my@tag@font\relax\amsmath@eqref{#1}}}
\makeatother

\allowdisplaybreaks

\begin{document}

\title{Systematic derivation of realistic spin-models for beyond-Heisenberg solids}

\newcommand{\fz}{Peter Gr\"unberg Institut and Institute for Advanced Simulation, Forschungszentrum J\"ulich and JARA, 52425 J\"ulich Germany}
\author{Markus Hoffmann}
\email{m.hoffmann@fz-juelich.de}  
\author{Stefan Bl\"ugel}
\affiliation{\fz}

\date{\today}

\begin{abstract}
We present a systematic derivation of effective lattice spin Hamiltonians derived from a rotationally invariant multi-orbital Hubbard model including a term ensuring Hund’s rule coupling. The Hamiltonians are derived down-folding the fermionic degrees of freedom of the Hubbard model into the proper low-energy spin sector using L\"owdin partitioning, which will be outlined in detail for the case of two sites and two orbitals at each site. Correcting the ground state systematically up to fourth order in the hopping of electrons, we find for spin $S\geq 1$ a biquadratic, three-spin and four-spin interaction beyond the conventional Heisenberg term. Comparing the puzzling energy spectrum of the magnetic states for a single Fe monolayer on Ru(0001), obtained from density functional theory, with the spin Hamiltonians taken at the limit of classical spins, we show that the previously ignored three-spin interaction can be comparable in size to the conventional Heisenberg exchange.

\end{abstract}

\maketitle

\section{Introduction}
Magnetic interactions have captivated several generations of condensed matter physicists because of their diversity of physical origins in very different solids, the emergence of a vast spectrum of magnetic structures as a result of their competition and subsequently the many interesting physical phenomena that are arising from those magnetic structures~\cite{Anderson:1963,Zeiger:1973,Graaf:2015}. Antiferromagnets with noncoplanar spin-textures and topological magnetization solitons such as skyrmions are current examples of complex magnetic structures with a broad spectrum of exotic properties that are of interest for both basic research and applications in spintronics~\cite{Fert:17}. Understanding the properties of these novel spin-textures has revitalized the field of magnetic interactions. In this context itinerant magnets play an important role as the itinerant electrons give rise to these complex magnetic structures and in turn the complex magnetic structures give rise to interesting transport phenomena~\cite{Nagaosa:01,Schultz:12,Diea:17}.  

In a materials specific context, the theoretical descriptions of magnetic ground states as well as the dynamical or thermodynamical properties of magnetic systems are often made possible by a realistic spin Hamiltonian typically determined by a multi-scale approach: density functional theory (DFT) calculations are mapped onto a classical lattice spin Hamiltonian, \textit{i.e.}\ a lattice of classical spins interacting according to spin-models, whose properties are then evaluated carrying out Monte-Carlo or spin-dynamic simulations~\cite{ncomms5030,PhysRevB.84.224413,PhysRevB.92.020401,gyorffy1985,PhysRevB.58.293,PhysRevB.64.174402,fahnle2007,PhysRevLett.107.017204,PhysRevLett.107.119901}. That is to say that the materials specificity enters through the parameters of the model determined by DFT. The choice of the spin-model itself reflects the choice of materials and the interactions that seem relevant to understand certain properties.  

For many bulk as well as application customized multilayer and heterostructure systems, the well-known spin ${S=1/2}$-Heisenberg model~\cite{Heisenberg:1928} of quantum spins $\op{S}$ is extrapolated to systems with higher quantum spin, ${S>1/2}$, and very often to classical vector spins $\vc{S}$ providing a parameterization of an effective spin Hamiltonian successful in describing the required magnetic properties. This holds also true for metallic magnetic materials, in particular those for which the longitudinal spin-fluctuations are unimportant as compared  to the transversal ones. These are typically magnets of transition metals with atomic spin moments in the order of 2~$\mu_\text{B}$ and more such as for Mn, Fe, Co in their bulk phases, as alloys and multilayers commonly used in spintronic devices. 

In fact, describing typical properties of those magnetic metals one resorts to the classical Heisenberg model of bilinear exchange interactions of the form 
\begin{equation}
H_1 = -\sideset{}{'}\sum_{ij} J_{ij}\, \vc{S}_i \cdot \vc{S}_j
\label{eq:Heisenberg}
\end{equation}
 between pairs of classical spins $\vc{S}$ at different lattice sites $i, j$ with exchange interactions $J_{ij}$ whose signs and strengths depend on details of the electronic structure. The spatial dependence of the exchange interaction follows typically the crystal anisotropy imposed by the crystal lattice. For metals the $J_{ij}$ can be long-ranged and in part determined by the topology of the Fermi surface, in opposite to insulators, where they are typically short-ranged. A success of this approach is for example the prediction of magnetic structures consistent to experiments~\cite{nature05802} or the Curie temperatures of bulk ferromagnets~\cite{doi:10.1080/14786430500504048,PhysRevB.72.184415}. The minus sign in \eqref{eq:Heisenberg} is just a convention we follow for all spin lattice Hamiltonians throughout the paper. The notation $\sum^\prime$ means here and throughout the paper that we are taking the sum over all possible integer sites $i$ and $j$ except for any summations of two equal sites $i=j$.
 
There are, however, well-known cases where the Heisenberg model is insufficient to describe correctly the magnetic ground state structure or magnon excitations. In these cases~\cite{3-note}
one addresses the higher-order spin interaction beyond the Heisenberg model. A typical signature of the higher-order spin interaction is the occurrence of particular types of non-collinear states, \textit{e.g.}\ canted magnetic states~\cite{Iwashita:1976} or multi-q states, a superposition of spin-spiral states of symmetry related wave vectors $q$. A spin-spiral state with a single q-vector~\cite{Iwashita:1976, PhysRevLett.101.027201} is an exact solution of the classical Heisenberg model for a periodic lattice. The higher-order terms couple modes of symmetry equivalent $q$ vectors and can lead to complex magnetic structures of energies lower than the single-q state~\cite{Kurz3q}.

One of the most commonly considered extensions of the bilinear Heisenberg form is the addition of the biquadratic exchange, a term of the form 
\begin{equation}
H_2 =- \sideset{}{'}\sum_{ij} B_{ij} (\vc{S}_i \cdot \vc{S}_j)^2\, .
\label{eq:biquadratic}
\end{equation}
This term has been motivated by very different microscopic origins, through superexchange~\cite{Anderson:1963}, magneto-elastic effect~\cite{Kittel:1960, Lines:1972} or interlayer exchange coupling~\cite{Bruno:1993}. Quite generally, according to the algebra of the spin operators, any power of scalar products of pairs of quantum spins  of total spin $S$ at sites $i$, $j$, can only have  $2S$ independent powers  up to $(\vop{S}_i \cdot \vop{S}_j)^{2S}$. Thus, for the biquadratic term to occur through the interaction of electrons requires  at least a  total spin ${S=1}$ at the lattice sites.  
As we will see below, as the power of $(\vop{S}_i \cdot \vop{S}_j)$  is related to the order of perturbation theory, the biquadratic term~\cite{0305-4470-10-3-011,PhysRevB.18.5078,PhysRev.140.A1803,Rodbell:1963,Gaulin:1986,Bhattacharjee:2006,Fridman:2000,Lou:2000,Brown:1975}
is the most essential correction to the Heisenberg model for spins ${S>1/2}$ involving two lattice sites. 

Involving more lattice sites, a systematic  extension of the bilinear Heisenberg form is the four-spin interaction, which was derived by Takahashi~\cite{takahashi} for a spin 1/2-system treating electrons by a single band Hubbard model. It arises  in fourth order perturbation theory of electron hopping versus Coulomb interaction~\cite{macdonald}. The four-spin interaction consists of four-body operators that appear by permuting all spins in a four-membered ring and can be written in the limit of classical spin as 
\begin{eqnarray}
H_4= -
\sideset{}{'}\sum_{ijkl} K_{ijkl}\,&&[(\mathbf{S}_i \cdot \mathbf{S}_j)(\mathbf{S}_k \cdot \mathbf{S}_l)+(\mathbf{S}_i \cdot \mathbf{S}_l)(\mathbf{S}_j \cdot \mathbf{S}_k)\nonumber \\
-&&(\mathbf{S}_i \cdot \mathbf{S}_k)(\mathbf{S}_j \cdot \mathbf{S}_l)]\, ,
\label{eq:4-spin}
\end{eqnarray}
with the sum over all rings of four sites.

\begin{figure*}[ht!]
    \centering
    \includegraphics[width=\textwidth]{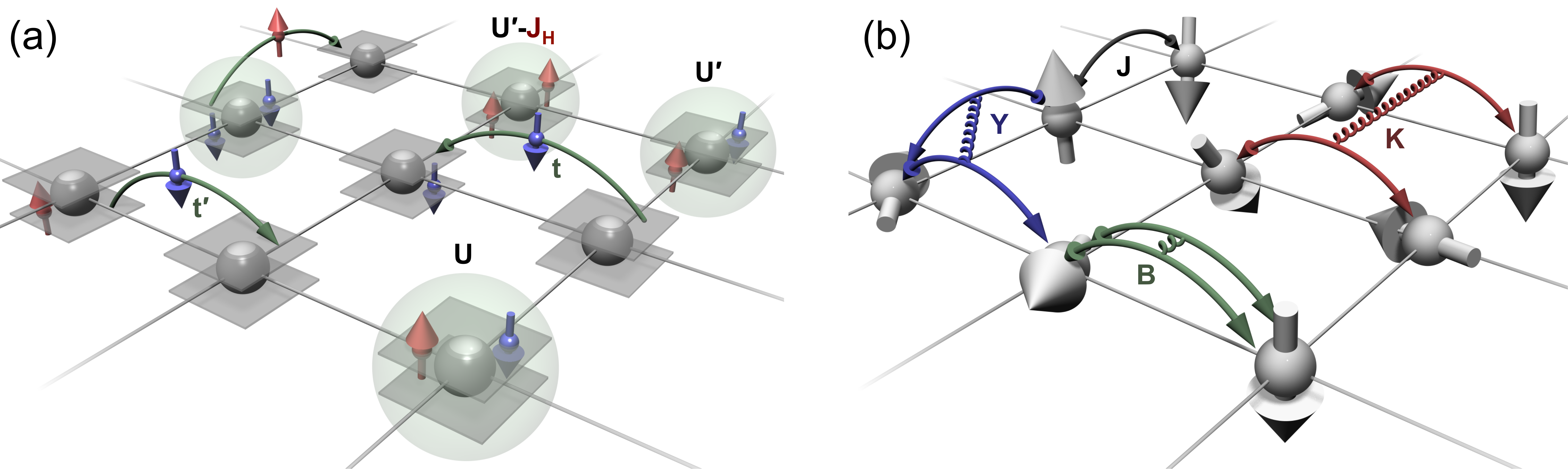}
    \caption{Schematic representation of the two investigated models. (a) The multi-band Hubbard model. A periodic arrangement of atoms on a lattice is shown. The different orbitals (here two) are illustrated by gray planes located at each atom. Each orbital can host up to two electrons, one spin-up (shown in red) and one spin-down (blue). Additionally, the hopping paths are indicated by green arrows and sites with non-zero on-site energies (proportional to $U$, $U'$ and $J_H$) are highlighted by green spheres. (b) The extended Heisenberg model in the limit of classical spins (gray arrows) at each lattice site. Direct exchange ($\vc{S}_i\cdot\vc{S}_j$) is illustrated by colored arrows. Higher-order interactions couple two of them to form 4-spin interactions involving two (B), three (Y) or four (K) sites as indicated by the springs.}
    \label{fig:schematic}
\end{figure*}

Although the higher-order spin models where mostly applied to magnets with localized electrons such as magnetic  insulators~\cite{Kobler1996,JPSJ.70.3089}, comparing DFT results for itinerant metallic magnets with spin-models reveals their significance also for these systems. Examples include contributions of the biquadratic term to the spin-stiffness of the bulk magnets Fe, Co and Ni~\cite{doi:10.1063/1.361678}, the conical spin spirals for a double-layer Mn on W(110)~\cite{silke}, or even three-dimensional non-collinear spin structures on a two-dimensional lattice as in Mn/Cu(111)~\cite{Kurz3q}, in Fe/Ir(100)~\cite{PhysRevB.92.020401} or
Fe/Ir(111)~\cite{heinze2011}. In case of the latter, the 4-spin interaction couples spin spirals with different propagation directions and forms
a square lattice of chiral magnetic skyrmions of atomic scale size.  

However, one became recently aware not all systems studied with  DFT could be explained purely on the basis of the higher-order interactions discussed above. Two such examples are the theoretically predicted \cite{PSSB:PSSB201147090} and recently experimentally verified~\cite{kronlein} so-called \textit{up-up-down-down} (uudd) state, a multi-q state, in Fe/Rh(111) or a canted uudd state in a RhFe bilayer system on Ir(111)~\cite{PhysRevLett.120.207201}. While an uudd state could in general be stabilized by both considered higher-order interactions independently, the calculated energy spectrum revealed that the main stabilization has to originate from another, hitherto unknown, interaction.

Summarizing the spin-models discussed so far we can view the Heisenberg, biquadratic and four-spin  model as a two-spin-two-site, four-spin-two-site, and four-spin-four-site interaction, respectively.
Heisenberg and four-spin interaction emerge for $S=1/2$, the biquadratic one requires at least $S=1$. Since typical magnetic Mn, Fe, Co moments
at surfaces are in the order of 2 or 3~$\mu_\text{B}$  equivalent to $S=1$ or $S=3/2$, there should be a large number of quasi two-dimensional non-Heisenberg magnets, in particular for substrates for which the effective Heisenberg exchange is small due to compensation of $J_{ij}$ of different signs between different neighbors.

Further, using this notion of classification, a four-spin-three-site interaction seems missing. Indeed, various partly phenomenological models of three-spin interactions~\cite{4-note}
had been proposed or derived to explain experiments mostly for insulating magnets~\cite{doi:10.1143/JPSJ.36.48,doi:10.1143/JPSJ.47.786,IWASHITA198064,19681169,PhysRev.140.A1803,Bastardis:2007,muller1997}.

In this paper we provide a consistent and systematic derivation of expressions describing the beyond-Heisenberg higher-order spin interactions resulting from the electron-electron interaction up to the fourth order in the hopping interaction strength of electrons for total spins of size $S\geq 1/2$. This includes all possible sequences of four hopping events of electrons between orbitals at maximal four sites.  
The spin-orbit interaction is neglected at this point. The starting point is the rotationally invariant multi-orbital Hubbard model assuming half-filling, which will be explained in the next section. The spin-model is derived down-folding the dynamical fermionic degrees of freedom of electrons described by the Hubbard model into the proper low-energy spin sector using L\"owdin partitioning~\cite{lowdin,winkler}, which is also known as Schrieffer-Wolf transformation~\cite{Schrieffer:66,Bravyi2011}. The L\"owdin partitioning is briefly sketched for a dimer of  $S=1$-spins described by two electron orbitals at both sites. Then, we will present our results for different numbers of sites and orbitals and also for  lattices with different space groups like a square lattice as for example for magnetic atoms on a (001)-surface of a fcc crystal, or on a hexagonal lattice like the (111)-surface to adapt the theoretical approach to real systems. Taking the classical spin-limit of the quantum spin-models derived, we reproduce the known spin Hamiltonians above plus the missing three-spin interaction
\begin{equation}
H_3= - 2\sideset{}{'}\sum_{ijk} Y_{ijk} (\mathbf{S}_i \cdot \mathbf{S}_j)(\mathbf{S}_j \cdot \mathbf{S}_k)\, ,
\label{eq:3-spin}
\end{equation}
where the sum goes over triangles of sites.
 
At the end, we will analyze the energy spectrum for various magnetic structures  determined by density functional theory for a single Fe monolayer on Ru(0001). Subsequently we will show that the hitherto puzzling results~\cite{PSSB:PSSB201147090} can finally be  understood.

\section{Methodology}

\subsection{Multi-band Hubbard model}
In this section we briefly introduce the Hamiltonian, from which we start our derivations, and define the most important parameters of our model. In the following section we will then focus on reducing the inherent degrees of freedom of the Hamiltonian to the spin degrees of freedom in order to derive effective spin models.  This Hamiltonian will then be used to generate spin Hamiltonians for  different systems that vary by the number of sites and orbitals and also by the lattice type.\par
Earlier similar investigations~\cite{takahashi, macdonald} typically used the one-band Hubbard model~\cite{hubbard,hubbard2,hubbard3} as a starting point since it is the simplest model for describing interacting electrons on a lattice.
For practical magnetic systems, which we have in mind with typical magnetic spin moments  on the order of 2 or 3~$\mu_\text{B}$ ($S=1$ or $S=3/2$), we extend our investigation to systems with more than one orbital per site (\textit{e.g.}\ $d$-orbitals of transition metals). Therefore, we work with a generalized Hubbard Hamiltonian, which not only includes the additional hopping terms and Coulomb interactions, but contains also additional terms to ensure Hund's rule coupling. The Hund's terms are included as we are interested in states with a fixed and stable magnetic moment $S$ per atom or site:
\begin{align}
\op{H} = & - \ \sum_{i<j,\alpha,\sigma} t_{i,\alpha,j,\alpha} \ \left( c_{i,\alpha,\sigma}^{\dagger}c_{j,\alpha,\sigma} + \mathrm{h.c.}\right) \nonumber\\
& - \ \sum_{\substack{i<j,\sigma\\ \alpha\neq\alpha'}} \ t'_{i,\alpha,j,\alpha'} \ (c_{i,\alpha,\sigma}^{\dagger}c_{j,\alpha',\sigma} + \mathrm{h.c.}) \nonumber\\
& + \ \sum_{i,\alpha} \ U_{i,\alpha} \ \hat{n}_{i,\alpha,\uparrow}\hat{n}_{i,\alpha,\downarrow}\nonumber\\
& + \ \sum_{\substack{i,\sigma\\\alpha<\alpha'}} U'_{i,\alpha,\alpha'} \ \left(\hat{n}_{i,\alpha,\sigma}\hat{n}_{i,\alpha',\sigma} + \hat{n}_{i,\alpha,\sigma}\hat{n}_{i,\alpha',\bar{\sigma}}\right)\nonumber\\
& - \ \sum_{\substack{i,\sigma\\\alpha<\alpha'}} J_{i,\alpha,\alpha'} \ \hat{n}_{i,\alpha,\sigma}\hat{n}_{i,\alpha',\sigma}\nonumber\\
& - \ \sum_{i,\alpha<\alpha'} J_{i,\alpha,\alpha'} \ \left( c_{i,\alpha,\uparrow}^{\dagger}c_{i,\alpha,\downarrow} c_{i,\alpha',\downarrow}^{\dagger}c_{i,\alpha',\uparrow} + \mathrm{h.c.} \right)\nonumber\\
& - \ \sum_{i,\alpha<\alpha'} J'_{i,\alpha,\alpha'} \ \left( c_{i,\alpha,\uparrow}^{\dagger}c_{i,\alpha',\uparrow} c_{i,\alpha,\downarrow}^{\dagger}c_{i,\alpha',\downarrow} + \mathrm{h.c.} \right)
\label{eq:Hamiltonian}\, .
\end{align}
Here, $i$ and $j$ represent the atomic sites, $\alpha$ and $\alpha'$ stand for the orbitals and $\sigma$ denotes the quantization of the spin projection of the electron ($\uparrow$ or $\downarrow$). $\hat{n}_{i,\alpha,\sigma} = c_{i,\alpha,\sigma}^{\dagger}c_{i,\alpha,\sigma}$ defines the number of electrons at site $i$ in orbital $\alpha$ with spin $\sigma$. $t$ ($t'$) describes the hopping amplitude between two different sites of the same (different~\cite{1-note-t2}) orbital types. The on-site hopping between different orbitals is not considered as we assume the orbitals to be orthogonal with respect to each other ($t'_{i,\alpha,i,\alpha'}=0$). Fig.~\ref{fig:schematic} shows a schematic visualization of the Hubbard as well as the effective spin model.

Only on-site Coulomb interactions are taken into account throughout the paper. Having a periodic solid in mind with only one atom type, we assume that the intra-orbital Coulomb interaction between electrons of the same orbitals $\alpha$ is the same for each site, $U_{i,\alpha}=U$, as well as the inter-orbital Coulomb interaction between electrons in different  orbitals, $U'_{i,\alpha,\alpha'}=U'$. Analogously, $J_{i,\alpha,\alpha'}=J_\text{H}$ and $J'_{i,\alpha,\alpha'}=J'_\text{H}$ simplifies due to the absence of the site dependency. \par

\subsection{L\"owdin partitioning}
 Here we briefly explain how L\"owdin partitioning~\cite{lowdin,winkler} is used to derive an effective spin Hamiltonian. As an example, we take the smallest interacting system with more than one orbital per site, two sites with two orbitals each. Assuming half-filled orbitals, we deal with four electrons, that could be distributed among the four available orbitals. Thus, an orbital, $\Ket{s\,}$, can be occupied with $s$ equal to one or two electrons or it can be unoccupied, denoted as $\Ket{\,\cdot\,}$. The possible states sorted according to the angular momentum quantum number $m$, representing the $z$-component of the total spin of the system include the following product states: 
\begin{align}
m=2: & \Ket{\uparrow,\uparrow,\uparrow,\uparrow} \nonumber \\
m=1: & \Ket{\uparrow,\uparrow,\uparrow,\downarrow},\Ket{\uparrow,\uparrow,\downarrow,\uparrow},\Ket{\uparrow,\downarrow,\uparrow,\uparrow},\Ket{\downarrow,\uparrow,\uparrow,\uparrow}, \nonumber\\
& \Ket{\uparrow\downarrow,\uparrow,\uparrow,\cdot},\Ket{\uparrow,\uparrow\downarrow,\uparrow,\cdot}, \Ket{\uparrow,\uparrow,\uparrow\downarrow,\cdot},\Ket{\uparrow\downarrow,\uparrow,\cdot,\uparrow}, \nonumber\\*
& \Ket{\uparrow,\uparrow\downarrow,\cdot,\uparrow},\Ket{\uparrow,\uparrow,\cdot,\uparrow\downarrow},\Ket{\uparrow\downarrow,\cdot,\uparrow,\uparrow},\Ket{\uparrow,\cdot,\uparrow\downarrow,\uparrow}, \nonumber\\*
& \Ket{\uparrow,\cdot,\uparrow,\uparrow\downarrow},\Ket{\cdot,\uparrow\downarrow,\uparrow,\uparrow},\Ket{\cdot,\uparrow,\uparrow\downarrow,\uparrow},\Ket{\cdot,\uparrow,\uparrow,\uparrow\downarrow} \nonumber\\
m=0: & \Ket{\uparrow,\uparrow,\downarrow,\downarrow}, \Ket{\uparrow,\downarrow,\uparrow,\downarrow}, \Ket{\uparrow,\downarrow,\downarrow,\uparrow}, \Ket{\downarrow,\uparrow,\uparrow,\downarrow}, \nonumber\\*
&  \Ket{\downarrow,\uparrow,\downarrow,\uparrow},\Ket{\downarrow,\downarrow,\uparrow,\uparrow}, \Ket{\uparrow\downarrow,\uparrow,\downarrow,\cdot}, \Ket{\uparrow\downarrow,\uparrow,\cdot,\downarrow}, \nonumber\\*
&  \Ket{\uparrow,\uparrow\downarrow,\downarrow,\cdot}, \Ket{\uparrow,\uparrow\downarrow,\cdot,\downarrow},\Ket{\uparrow\downarrow,\downarrow,\uparrow,\cdot}, \Ket{\uparrow\downarrow,\cdot,\uparrow,\downarrow}, \nonumber\\*
&  \Ket{\uparrow,\downarrow,\uparrow\downarrow,\cdot}, \Ket{\uparrow,\cdot,\uparrow\downarrow,\downarrow}, \Ket{\uparrow\downarrow,\downarrow,\cdot,\uparrow},\Ket{\uparrow\downarrow,\cdot,\downarrow,\uparrow}, \nonumber\\*
&  \Ket{\uparrow,\downarrow,\cdot,\uparrow\downarrow}, \Ket{\uparrow,\cdot,\downarrow,\uparrow\downarrow}, \Ket{\downarrow,\uparrow\downarrow,\uparrow,\cdot}, \Ket{\downarrow,\uparrow,\uparrow\downarrow,\cdot}, \nonumber\\* 
&  \Ket{\cdot,\uparrow\downarrow,\uparrow,\downarrow}, \Ket{\cdot,\uparrow,\uparrow\downarrow,\downarrow}, \Ket{\downarrow,\uparrow\downarrow,\cdot,\uparrow}, \Ket{\downarrow,\uparrow,\cdot,\uparrow\downarrow}, \nonumber\\*
&  \Ket{\cdot,\uparrow\downarrow,\downarrow,\uparrow},\Ket{\cdot,\uparrow,\downarrow,\uparrow\downarrow}, \Ket{\downarrow,\cdot,\uparrow\downarrow,\uparrow}, \Ket{\downarrow,\cdot,\uparrow,\uparrow\downarrow}, \nonumber\\*
&  \Ket{\cdot,\downarrow,\uparrow\downarrow,\uparrow}, \Ket{\cdot,\downarrow,\uparrow,\uparrow\downarrow}, \Ket{\uparrow\downarrow,\uparrow\downarrow,\cdot,\cdot}, \Ket{\uparrow\downarrow,\cdot,\uparrow\downarrow,\cdot}, \nonumber\\*
&  \Ket{\uparrow\downarrow,\cdot,\cdot,\uparrow\downarrow}, \Ket{\cdot,\uparrow\downarrow,\uparrow\downarrow,\cdot}, \Ket{\cdot,\uparrow\downarrow,\cdot,\uparrow\downarrow},\Ket{\cdot,\cdot,\uparrow\downarrow,\uparrow\downarrow} \nonumber\\
m=-1: &  \Ket{\uparrow,\downarrow,\downarrow,\downarrow}, \Ket{\downarrow,\uparrow,\downarrow,\downarrow}, \Ket{\downarrow,\downarrow,\uparrow,\downarrow}, \Ket{\downarrow,\downarrow,\downarrow,\uparrow}, \nonumber\\*
&  \Ket{\uparrow\downarrow,\downarrow,\downarrow,\cdot},\Ket{\uparrow\downarrow,\downarrow,\cdot,\downarrow}, \Ket{\uparrow\downarrow,\cdot,\downarrow,\downarrow}, \Ket{\downarrow,\uparrow\downarrow,\downarrow,\cdot}, \nonumber\\*
&  \Ket{\downarrow,\uparrow\downarrow,\cdot,\downarrow}, \Ket{\cdot,\uparrow\downarrow,\downarrow,\downarrow}, \Ket{\downarrow,\downarrow,\uparrow\downarrow,\cdot}, \Ket{\downarrow,\cdot,\uparrow\downarrow,\downarrow}, \nonumber\\*
&  \Ket{\cdot,\downarrow,\uparrow\downarrow,\downarrow}, \Ket{\downarrow,\downarrow,\cdot,\uparrow\downarrow}, \Ket{\downarrow,\cdot,\downarrow,\uparrow\downarrow},\Ket{\cdot,\downarrow,\downarrow,\uparrow\downarrow} \nonumber\\
m=-2: &  \Ket{\downarrow,\downarrow,\downarrow,\downarrow}
\end{align}
Here, $\Ket{s_1, s_2, s_3, s_4} = \Ket{s_1}\Ket{s_2}\Ket{s_3}\Ket{s_4}$ means that at site 1 the first (second) orbital is occupied by $s_1$ ($s_2$) and at site 2 the first (second) orbital is occupied by $s_3$ ($s_4$). In general, for a  system with $n$ orbitals, the number of states for each value of $m$ is given by $\binom{n}{n/2 + m}^2$. 

Since the z-component of the angular momentum vector operator $\op{S}_z$ commutes with the Hamiltonian~\eqref{eq:Hamiltonian}, the Hamiltonian block-diagonalizes in separate subspaces of different $m$, and the matrix representation of \eqref{eq:Hamiltonian} can be calculated for each subspace separately. To support our goal of contracting Hamiltonian \eqref{eq:Hamiltonian} of our model to an effective spin Hamiltonian, it is convenient to change the product basis $\Ket{s_1, s_2, s_3, s_4}$ to one where the total spin at any site is a good quantum number. For example, for  $m=1$ the first 4 states are replaced by the following superpositions:
\begin{equation}
\begin{split}
\frac{1}{\sqrt{2}} \left( \Ket{\uparrow, \uparrow, \uparrow, \downarrow} + \Ket{\uparrow, \uparrow, \downarrow, \uparrow} \right) & = \Ket{1, 1} \Ket{ 1,0}\\
 \frac{1}{\sqrt{2}} \left( \Ket{\uparrow, \downarrow, \uparrow, \uparrow} + \Ket{\downarrow, \uparrow, \uparrow, \uparrow} \right) & = \Ket{1, 0} \Ket{ 1,1}\\
\frac{1}{\sqrt{2}} \left( \Ket{\uparrow, \uparrow, \uparrow, \downarrow} - \Ket{\uparrow, \uparrow, \downarrow, \uparrow} \right) & = \Ket{1, 1} \Ket{ 0,0}\\
\frac{1}{\sqrt{2}} \left( \Ket{\uparrow, \downarrow, \uparrow, \uparrow} - \Ket{\downarrow, \uparrow, \uparrow, \uparrow} \right) & = \Ket{0, 0} \Ket{ 1,1}\, ,
\label{eq:spinbasis}
\end{split}
\end{equation}
where we used the notation $\Ket{S_1, m_1} \Ket{S_2,m_2}$ with $S_i$ being the spin quantum number and $m_i$ being the total z-component at site $i$.\par
We are essentially interested in the subspace spanned by the first two states of \eqref{eq:spinbasis} as we assume magnetic systems, which have constant magnetic moments (here, $S=1$) at each site. Although there is no direct interaction between these two states, there are indirect interactions across states where $S$ is not equal at all sites. These indirect interactions between intermediate states in different subspaces can be downfolded into the sector of interacting spins of constant quantum number at each site using the so-called L\"owdin partitioning~\cite{lowdin,winkler}. L\"owdin partitioning can be used because we are dealing with energetically well separated subspaces of spins with different $S$. This is a consequence of Hund coupling and on-site Coulomb energies that are large with respect to the hopping parameters, as we are discussing transition metals here.\par
The L\"owdin partitioning is a tool to decouple these subspaces pertubatively and to map the indirect interaction between two states of the same subspace over states of the other subspaces to direct interactions between these states with increasing order of the perturbation. \Eg the indirect interaction
 $\Ket{\uparrow,\uparrow,\downarrow,\downarrow} \xleftrightarrow{\sim t} \Ket{\uparrow,\cdot,\downarrow,\uparrow\downarrow} \xleftrightarrow{\sim t} \Ket{\uparrow,\downarrow,\downarrow,\uparrow}$
is mapped on a direct interaction
$ \Ket{\uparrow,\uparrow,\downarrow,\downarrow} \xleftrightarrow{\sim t^2/U} \Ket{\uparrow,\downarrow,\downarrow,\uparrow}$
if terms up to at least second order are taken into account in the L\"owdin partitioning. By going  to higher orders also indirect interactions including more than two 
hopping events are considered. These can then relate to interactions with more than two sites.\par
Mathematically, this is achieved by dividing the Hamiltonian $\mathcal{H}$ into two parts, 
\begin{equation}
     \mathcal{H} = \mathcal{H}_0 + \mathcal{H}' = \mathcal{H}_0 + \mathcal{H}_1 + \mathcal{H}_2\, ,
\end{equation}
a term $\mathcal{H}_0$ that contains the on-site contributions, \textit{i.e.}\ the repulsive Coulomb interaction and the Hund exchange,   and a term $\mathcal{H}'$, which contains  the off-diagonal matrix elements due to the electron hopping, which are treated as a perturbation. Here, $\mathcal{H}_1$ contains those terms whose matrix elements couple within the subspaces, whereas $\mathcal{H}_2$ describes the coupling between them. The subspaces are decoupled  through a canonical transformation~\cite{Schrieffer:66,Bravyi2011}
\begin{equation}
  \tilde{\mathcal{H}} = e^{-\mathcal{\hat{S}}}\mathcal{H}e^{\mathcal{\hat{S}}}\, ,
\label{eq:canonical}
\end{equation}
where hermiticity of the Hamiltonian implies $\mathcal{\hat{S}}^\dagger = -\mathcal{\hat{S}}$ and the generator $\mathcal{\hat{S}}$ of the transformation is chosen such that $\tilde{\mathcal{H}}$ becomes block-diagonal. This is achieved writing Eq.~\eqref{eq:canonical} in the form of successive applications of commutator rules
\begin{equation}
\tilde{\mathcal{H}} =  \sum_{k=0}^{\infty} \frac{1}{k!}\left[ \mathcal{H},\mathcal{\hat{S}}\right]^k =  \sum_{k=0}^{\infty} \frac{1}{k!}\left[ \mathcal{H}_0+\mathcal{H}_1+\mathcal{H}_2,\mathcal{\hat{S}}\right]^k \, ,
\label{eq:sum} 
\end{equation}
with $\commutator{A}{B}^{k}=\commutator{\commutator{\ldots \commutator{\commutator{A}{B}}{B}\ldots}{B}}{B}$ nested $k$ times.
Considering the definitions of $\mathcal{H}_1$ and $\mathcal{H}_2$~\cite{winkler}, this allows then 
to decouple Eq.~\eqref{eq:sum} into a  Hamiltonian term $\tilde{\mathcal{H}}_\text{d}$, whose matrix representation is block diagonal and a term $\tilde{\mathcal{H}}_\text{o}$ with off-block-diagonal matrix elements as shown here:
\begin{equation}
\begin{split}
 \tilde{\mathcal{H}}_\text{d} = & \sum_{k=0}^{\infty} \frac{1}{(2k)!}\left[ \mathcal{H}_0\!+\!\mathcal{H}_1,\mathcal{\hat{S}}\right]^{2k} + \sum_{k=0}^{\infty} \frac{1}{(2k+1)!}\left[ \mathcal{H}_2,\mathcal{\hat{S}}\right]^{2k+1}\\
 \tilde{\mathcal{H}}_\text{o} = & \sum_{k=0}^{\infty} \frac{1}{(2k+1)!}\left[ \mathcal{H}_0\!+\!\mathcal{H}_1,\mathcal{\hat{S}}\right]^{2k+1} + \sum_{k=0}^{\infty} \frac{1}{(2k)!}\left[ \mathcal{H}_2,\mathcal{\hat{S}}\right]^{2k}
\end{split}
\label{eq:split}
\end{equation}
The requirement of block diagonalization or  $\tilde{\mathcal{H}}_\text{o}=0$, respectively, up to a given order $k$ in the perturbation determines the generator $\mathcal{\hat{S}}$ and subsequently the effective Hamiltonian $\tilde{\mathcal{H}}_\text{d}$. Due to the block-diagonalization of  $\mathcal{H}$ with respect to the basis of  $\vop{S}_z$, the L\"owdin partitioning can be carried out independently for  each angular momentum quantum number $m$. We work out all spin models for either $m=0$ or $m\pm 1/2$, depending the systems have integer or half-integer total spins, since these states denote the largest subspaces, and the L\"owdin partitioning becomes least degenerate and the functional forms of the spin Hamiltonians become most obviously distinct.  

\section{Results}
\subsection{Derived spin Hamiltonians}

Recalling that the spin operators 
\begin{equation}
 \vop{S}_i = \left( \op{S}_{i,x}, \op{S}_{i,y}, \op{S}_{i,z}\right)
 \label{eq:Si}
\end{equation}
can be expressed by the electron operators $c_{i,\alpha,\sigma}$, $c_{i,\alpha,\sigma}^{\dagger}$ as
\begin{align}
 \op{S}_{i,x}\ =\ & \phantom{-i}\frac{1}{2}\sum_{\alpha}\left(c_{i,\alpha,\uparrow}^{\dagger}c_{i,\alpha,\downarrow} + c_{i,\alpha,\downarrow}^{\dagger}c_{i,\alpha,\uparrow}\right) \nonumber \\
 \label{eq:S-elec}
 \op{S}_{i,y}\ =\ & -\frac{i}{2}\sum_{\alpha}\left(c_{i,\alpha,\uparrow}^{\dagger}c_{i,\alpha,\downarrow} - c_{i,\alpha,\downarrow}^{\dagger}c_{i,\alpha,\uparrow}\right) \\
 \op{S}_{i,z}\ =\ &\phantom{-i} \frac{1}{2}\sum_{\alpha}\left(\hat{n}_{i,\alpha,\uparrow}-\hat{n}_{i,\alpha,\downarrow}\right) \, ,\nonumber
\end{align}
whereby the sum goes over all orbitals $\alpha$ at site $i$, we show now how the electron Hamiltonian of a particular model system  folded down in the proper spin sector can be expressed by spin operators and thus represents the corresponding spin Hamiltonian or spin model of the system. In the following we present results up to fourth-order perturbation in \eqref{eq:split} which permits the investigation of interactions between 2, 3, and 4 sites. We start with spin $S=1/2$, \textit{i.e.}\ exactly one orbital per site, and then move to $S\ge 1$. \par

\subsubsection{Spin S=1/2}
\paragraph{2 sites, spin $S=1/2$}
To demonstrate the general procedure, we first discuss a $S=1/2$ dimer, \textit{i.e.}, two sites and only one orbital per site.  
For $m=0$, the Hilbert space is spanned by four possible states 
 $\Ket{\uparrow,\downarrow}, \Ket{\downarrow,\uparrow}, \Ket{\uparrow\downarrow,\cdot}, \Ket{\cdot,\uparrow\downarrow}$,
from which the first two span the subspace of interest with $S=1/2$ at both sites. $\mathcal{H}_0$ gives the same on-site energies for both states, which we consider the origin  of our energy scale. 
Going up to second order in the perturbation (the first order vanishes, because there is no direct coupling between the states) additional terms occur which couple the states. Those terms are, \textit{e.g.}\ proportional to
 $c_{2,\uparrow}^{\dagger}c_{1,\downarrow}^{\dagger}c_{1,\uparrow}c_{2,\downarrow} + h.c.$,
representing a hopping
$ \Ket{\uparrow, \downarrow} \leftrightarrow \Ket{\downarrow, \uparrow}$.
Collecting all those terms and extending the derivation to an infinite lattice of two-site interactions,  the resulting Hamiltonian can be written in terms of the spin operators \eqref{eq:S-elec} as
\begin{equation}
\begin{split}
\mathcal{H}^{\text{2 sites}}_{2^\text{nd}\text{order}}
 &= \frac{4\, t^2}{U} \sideset{}{'}\sum_{ij}c_{i,\uparrow}^{\dagger}c_{i,\downarrow}c_{j,\downarrow}^{\dagger}c_{j,\uparrow} - \hat{n}_{i,\uparrow}\hat{n}_{j,\downarrow}\\ 
&= \frac{2\, t^2}{U} \sideset{}{'}\sum_{ij}\left( \vop{S}_i \cdot \vop{S}_j - \frac{\hat{n}_i \hat{n}_j}{4}\right) \, , 
\end{split}
\label{eq:heis}
\end{equation}
with $\hat{n}_{i} = \left(\hat{n}_{i,\uparrow}+\hat{n}_{i,\downarrow}\right)$ being the total number operator with expectation value $n_i$ for electrons at site $i$. As we only consider the low-energy subspace, charge excitations are neglected, and only states with half-filled orbitals giving rise to maximal $S$ are considered. Thus, the last term in Eq.~\eqref{eq:heis}
defines a constant energy shift by ${n_1 n_2}/{4} = |S_1|\cdot |S_2| = S^2$~\cite{2-note-ferro}.
Thus, by going just up to second order in the perturbation of the hopping terms we obtain the well-known Heisenberg term \eqref{eq:Heisenberg}, if we define the exchange parameter $J$ as $J=-2 t^2/U$.\par

According to what has been said above, $S=1/2$ models with pair-interaction involving electrons hopping between two sites can only exhibit a bilinear spin Hamiltonian. This is confirmed by the inclusion of fourth order terms in the perturbation \eqref{eq:sum} (the third order vanishes again), which can be summarized  to the following expression
\begin{equation}
\mathcal{H}^{\text{2 sites}}_{4^\text{th} \text{order}} =-\frac{8\, t^4}{U^3} \sideset{}{'}\sum_{ij}\left( \vop{S}_i \cdot \vop{S}_j - \frac{\hat{n}_i \hat{n}_j}{4}\right) \, .
\label{eq:heis4}
\end{equation}
No terms of additional spin-spin interactions show up in fourth order perturbation for two  site-interactions. This shows that indeed a system of pair interactions of spin-1/2 sites  can be described purely by the Heisenberg interaction \eqref{eq:Heisenberg}, although the fourth-order term provides a correction of the Heisenberg exchange parameter
\begin{equation}
 J = - \frac{2\, t^2}{U} + \frac{8t^4}{U^3}\, .
\end{equation}
The negative sign of the leading term means that the magnetic interaction of a spin-1/2 system is the $m=0$ singlet state if $t/U<1/2$, which we equate with the antiferromagnetic state. If the system becomes more metallic, the hopping matrix element $t$ increases as well as the number of sites involved. Then the prefactor of the second term increases rapidly with system size (see Tab.~\ref{tab:hubbard}) and the likelihood for ferromagnetic interactions increases. Although discussed only for spin states with $m=0$, the same effective Hamiltonian is also able to describe those with $m=1$ and $m=-1$, respectively, as the excitation energy of those states due to the hopping of electrons is $0$ both in the Hubbard Hamiltonian and in the effective spin Hamiltonian.
\par
\paragraph{3 sites, spin $S=1/2$}
  Since the fourth-order perturbation term in \eqref{eq:split} involves four successive hopping events of electrons, the interaction can involve spins or orbitals, respectively, beyond two sites up to four sites and thus can go beyond the pair interaction typical for the Heisenberg model.  Considering three sites, the perturbation theory results, however, again in a pair interaction analogous to the Heisenberg model. The only difference with respect to the system with two sites is a change of the prefactor, \ie of the Heisenberg exchange parameter, respectively (\textit{cf.}\ Tab.~\ref{tab:hubbard}, for simplicity we assumed the same $t$ for all hopping events. The effect of different hopping elements $t_{ij}$ will be analyzed below). Again, this spin Hamiltonian is capable of describing all the subspaces for different $m$ (here, $m = {-3/2, -1/2, +1/2, +3/2}$).\par
\paragraph{4 sites, spin $S=1/2$}
For four sites, the fourth order perturbation produces terms, which can be subsumed to the Heisenberg term, but generates also additional ones, for example, 
$ c_{4,\uparrow}^{\dagger}c_{3,\uparrow}^{\dagger}c_{2,\downarrow}^{\dagger}c_{1,\downarrow}^{\dagger}c_{1,\uparrow}c_{2,\uparrow}c_{3,\downarrow}c_{4,\downarrow} + h.c. $, 
 for four sites with $m=0$. In contrast to the terms above, this term flips four spins instead of two. \par
If we collect all these terms of fourth order and express them in terms of spin operators we obtain 
\begin{equation}
 \mathcal{H}^{\text{4 sites}}_{4^\text{th} \text{order}} =\frac{10\,t^4}{U^3} \sideset{}{'}\sum_{ijkl} \left(\vop{S}_i\cdot \vop{S}_j - \frac{n_i n_j}{4}\right)\left(\vop{S}_k\cdot \vop{S}_l - \frac{n_k n_l}{4}\right)\, ,
 \label{eq:4-sites}
\end{equation}
which can be divided into a 4-spin term
\begin{equation}
 \mathcal{H}^{\text{4 sites}}_{\text{4 spins}} =\frac{10\,t^4}{U^3} \sideset{}{'}\sum_{ijkl} \left(\vop{S}_i\cdot \vop{S}_j\right)\left(\vop{S}_k\cdot \vop{S}_l\right)\, ,
\label{eq:4th-order}
\end{equation}
plus a Heisenberg term with the prefactor $J={10\,t^4}/{U^3}$, and a constant energy shift of size ${15\,t^4}/{U^3}$. The prefactor  in Eq.~\eqref{eq:4th-order} will be called $-K$ in this paper.\par

\begin{table}[t]
\caption{Calculated prefactors of the Heisenberg exchange and the 4-spin interactions in terms of the model parameters $t$ and $U$ of the Hubbard model \eqref{eq:Hamiltonian} taken at single-orbital per site for different numbers of sites with $S=1/2$ obtained by going up to 4$^\text{th}$ order in the L\"owdin partitioning.}
\label{tab:hubbard}
 \begin{tabular}{c || c | c}
 \hline\hline
 & & \\[-7pt]
sites & $J$ & $K$\\[2pt]
\hline
  & &  \\[-7pt]
2 & $-\frac{2\, t^2}{U} + \frac{8\, t^4}{U^3}$ & $0$\\[5pt]
3 & $-\frac{2\, t^2}{U} + \frac{6\, t^4}{U^3}$ & $0$\\[5pt]
4 & $-\frac{2\, t^2}{U} + \frac{10\, t^4}{U^3}$ & $-\frac{10\, t^4}{U^3}$\\[5pt]
5 & $-\frac{2\, t^2}{U} + \frac{20\, t^4}{U^3}$ & $-\frac{10\, t^4}{U^3}$\\[5pt]
6 & $-\frac{2\, t^2}{U} + \frac{36\, t^4}{U^3}$ & $-\frac{10\, t^4}{U^3}$\\[5pt]
8 & $-\frac{2\, t^2}{U} + \frac{86\, t^4}{U^3}$ & $-\frac{10\, t^4}{U^3}$
\end{tabular}
\end{table}
\ \par

Equation \eqref{eq:4th-order} is a simplified version of the more complex four-spin interaction~\cite{ncomms5030, PhysRevB.92.020401, heinze2011} introduced in  \eqref{eq:4-spin}, namely for the case when the hopping parameters between all the atoms are the same. In a real system this is rarely the case as the value of the hopping parameter $t$ depends  on the distances between the two involved atoms, the types of orbitals, but also on the environment, for details see also Section:~\ref{sec:surfaces}. Carrying out a  more explicit calculation of the fourth-order term  with pair-dependent hopping parameter $t_{ij}$, the prefactor $K\propto -t^4$ in \eqref{eq:4-sites}, \eqref{eq:4th-order} changes to ring paths of hopping  with $K_{ijkl} \propto - t_{ij}t_{jk}t_{kl}t_{li}$ and with spin terms as in  \eqref{eq:4-spin}.

\paragraph{$N>4$ sites, spin $S=1/2$}
Going up to more sites (\eg 5, 6, and 8) 
we showed no additional spin interaction terms emerge and the previously shown spin Hamiltonians (Heisenberg plus 4-spin) describe fully the energy landscape. The calculated prefactors for the case that the same hopping parameter $t$ exists between all sites are shown in Tab.~\ref{tab:hubbard}. Additional interaction terms will emerge beyond fourth order perturbation calculations, \eg six-order terms for $N\ge 6$, which is beyond the scope of this paper.

\subsubsection{Spin S $\ge$ 1}

The extension to systems with larger spins per site, which is made possible by more than one half-filled orbital per site, is in principle straightforward, but in practice significantly more complex. The Hilbert space becomes much larger and we need to switch from a single-band to a multi-band Hubbard model with quite some additional interaction parameters, which ultimately adds considerable complexity to the  prefactors or the exchange parameters of the spin models, respectively (see Appendix for details).  To keep the prefactors simple and transparent,  we discuss here results for the simplified case, where hopping interactions between equal and different orbitals are identical and orbital independent, $t'=t$, and the Coulomb repulsion and the exchange interaction of electrons at the same site but different orbitals, $U'=0$ and $J'=0$, are neglected (see Tab.~\ref{tab:JBYK}), valid assuming   that the Coulomb energy is larger if the electrons are not just at the same site but also in the same orbital, \textit{i.e.}\ for $U'\ll U$ and $J'\ll J$. However, these simplifications do not alter the functional nature of the spin models, just simplify prefactors. The full prefactors can be found in the appendix.\par

\paragraph{2 sites, spin $S=1$} Starting again with the simplest $S=1$ model of two sites with two orbitals per site, we find in second order perturbation terms in which two spins are reversed. For example, for $m=0$
 $c_{2,2,\downarrow}^{\dagger}c_{1,1,\downarrow}^{\dagger}c_{1,1,\uparrow}c_{2,2,\uparrow}$. 
is such a term.
As it can be seen, there is always one orbital per site involved in those spin flips. Collecting now all the terms which arise in second order perturbation we again end up with the Heisenberg Hamiltonian , but with a prefactor of $J = -2t^2/(U+J_\text{H})$. \par
In fourth order perturbation, however, more important differences to the system with only one orbital per site occur. In addition to the term shown above, there appear additional terms as 
\begin{equation}
 c_{2,2,\downarrow}^{\dagger}c_{2,1,\uparrow}^{\dagger}c_{1,2,\uparrow}^{\dagger}c_{1,1,\downarrow}^{\dagger}c_{1,1,\uparrow}c_{1,2,\downarrow}c_{2,1,\downarrow}c_{2,2,\uparrow}\, ,
\end{equation}
where all the electron spins are reversed and thus all orbitals are involved in this interaction. For this reason, we can already see that each site is involved twice in this interaction and therefore has to occur twice in the effective spin Hamiltonian. And indeed, by using the spin operators, the resulting effective interaction can be written as
\begin{equation}
\mathcal{H}^{\text{2 sites}}_{4^\text{th}\text{order}}\propto
  \sideset{}{'}\sum_{ij}\left( \vop{S}_i \cdot \vop{S}_j - \frac{\hat{n}_i \hat{n}_j}{4}\right)^2\, ,
\end{equation}
which can be simplified into the biquadratic interaction \eqref{eq:biquadratic},
a Heisenberg term, and a constant energy shift. The prefactors for this system and the systems introduced in the following with the previously named assumptions on the parameters of the multi-band Hamiltonian \eqref{eq:Hamiltonian} can be found in Tab.~\ref{tab:JBYK}. The prefactors for systems treated with an unrestricted parameter set are shown in the Appendix.\par
\begin{table}[t]
\caption{Calculated prefactors of the Heisenberg exchange ($J$), the biquadratic ($B$), the three-spin ($Y$) and the four-spin interactions ($K$) for different numbers of sites with $S>1/2$ obtained by going up to 4$^\text{th}$ order in the L\"owdin partitioning. We set $t'=t$ and $U'$ and $J'$ were set to zero (see text).}
\label{tab:JBYK}
 \begin{tabular}{c c || c | c | c | c}
 \hline\hline
  & & & & & \\[-7pt]
sites & S & $J$ & $B$ & $Y$ & $K$\\[2pt]
\hline
 & & & & & \\[-7pt]
2 & 1 & $-\frac{2t^2}{U+J_\text{H}}$ & $\frac{-20t^4}{(U+J_\text{H})^3}$ & 0 & 0\\[5pt]
3 & 1 & $-\frac{2t^2}{U+J_\text{H}} + \frac{36t^4}{(U+J_\text{H})^3}$ & $\frac{-20t^4}{(U+J_\text{H})^3}$ & $\frac{-40t^4}{(U+J_\text{H})^3}$ & 0\\[5pt]
4 & 1& $-\frac{2t^2}{U+J_\text{H}} + \frac{96t^4}{(U+J_\text{H})^3}$ & $\frac{-20t^4}{(U+J_\text{H})^3}$ & $\frac{-40t^4}{(U+J_\text{H})^3}$ & $\frac{-10t^4}{(U+J_\text{H})^3}$ \\[5pt]
2 & 3/2& $-\frac{2t^2}{U+J_\text{H}} + \frac{6t^4}{(U+J_\text{H})^3}$ & $\frac{-20t^4}{(U+J_\text{H})^3}$ & 0 & 0 \\[5pt]
3 & 3/2& $-\frac{2t^2}{U+J_\text{H}} + \frac{90t^4}{(U+J_\text{H})^3}$ & $\frac{-20t^4}{(U+J_\text{H})^3}$ & $\frac{-40t^4}{(U+J_\text{H})^3}$ & 0 \\[4pt]
\end{tabular}
\end{table}
The appearance of the biquadratic interaction for $S=1$ dimers is consistent with the spin-algebra, which states that the highest independent powers of pair interactions is given by $(\vop{S}_i \cdot \vop{S}_j)^{2S}$. For $S=1/2$ dimers, the biquadratic term can always be expressed as the sum of the Heisenberg term and a constant shift, and thus disappears. Similarly, in $S=1$ systems higher powers of $(S_1 \cdot S_2)^n$, with $n \ge 3$, can be expressed in a sum of the biquadratic and  Heisenberg term as well as a constant shift, and disappear too.

\paragraph{3 sites, spin $S=1$} 
Considering a system with three sites and two orbitals at each site, second order perturbation theory reproduces again the Heisenberg model between different pairs of the three sites. Fourth order perturbation enables the reverse of spins in four different orbitals, which in a system with 3 sites can be facilitated in two different ways: Either the four orbitals are taken just at two different sites or they are distributed over all three sites, of which one is the site where the electron spin is reversed in both orbitals, while at each of  the other two sites only one orbital is involved in the hopping. The former one results again in a biquadratic interaction. The latter one includes terms like
\begin{equation}
 \hat{n}_{3,1,\uparrow}\hat{n}_{1,2,\downarrow}c_{3,2,\downarrow}^{\dagger}c_{2,2,\uparrow}^{\dagger}c_{2,1,\uparrow}^{\dagger}c_{1,1,\downarrow}^{\dagger}c_{1,1,\uparrow}c_{2,1,\downarrow}c_{2,2,\downarrow}c_{3,2,\uparrow} \ \ \ ,
\end{equation}
where we can clearly see that two orbitals (here, the first orbital at site 3 and the second at site 1) are not affected by this hopping term, while the other four change their spin direction. At the end this can be summarized in terms of an effective Hamiltonian
\begin{equation}
\mathcal{H}^{\text{3 sites}}_{4^\text{th}\text{order}}\propto
\sideset{}{'}\sum_{ijk} \left(\vop{S}_i\cdot \vop{S}_j - \frac{\hat{n}_i \hat{n}_j}{4}\right)\left(\vop{S}_i\cdot \vop{S}_k - \frac{\hat{n}_i \hat{n}_k}{4}\right)\, ,
\end{equation}
 which again can be structured into three different terms namely a Heisenberg term, a constant shift and
\begin{equation}
\mathcal{H}^{\text{3 sites}}_{\text{4 spins}}\propto
\sideset{}{'}\sum_{ijk} \left(\vop{S}_i\cdot \vop{S}_j\right)\left(\vop{S}_i\cdot \vop{S}_k\right)\ \ \ ,
\label{eq:3spin-derived}
\end{equation}
an Hamiltonian expression we identify as the three-spin interaction introduced in \eqref{eq:3-spin}. The exchange constant of the 3-spin interaction is called $Y$ henceforth. As we can see from the collection of prefactors in Tab.~\ref{tab:JBYK} and Appendix, the 3-spin constant, $Y$, is in the same order of magnitude and even by a factor of 2 larger than the biquadratic constant, $B$. Therefore, we suppose that this 3-spin interaction can play an important role in systems in which other higher order interactions such as the biquadratic or 4-spin interaction are comparable in size to the Heisenberg one. Iron based thin-film systems are candidates for such a behavior, because the local magnetic moments and spins, respectively, of Fe are large in these environments. We will demonstrate this below for the exemplary systems Fe/Rh(111) and Fe/Ru(0001).\par
\paragraph{4 sites, spin $S=1$}
The behavior within the second order perturbation is the same as before. However, within fourth order perturbation calculations  and in comparison to the derivation of the interaction across three sites additional interaction terms are expected since the four orbitals which are involved in the interactions of the fourth order perturbation can now either be divided-up over 2, 3 or 4 sites resulting in the biquadratic, 3-spin and 4-spin interaction, respectively. The prefactors can be found in Tab.~\ref{tab:JBYK}. So we have shown that the 4-spin interaction is not just a result that occurs in $S=1/2$ systems, but also in those with $S=1$.\par
\paragraph{spin $S> 1$} To clarify whether the previously shown results apply only to $S=1$-systems or can also be applied to systems with larger spins, we have also investigated systems with $S=3/2$ that represent systems having three orbitals per site exhibiting local magnetic moments of 3~$\mu_\text{B}$. As we can see in Tab.~\ref{tab:JBYK} the considered systems can all be explained by the interplay of the exchange, biquadratic, 3-spin and 4-spin interaction.\par
Additional magnetic interaction terms making use of the nature of at least three orbitals per site would require the concerted hopping of six electrons, which is beyond the fourth order perturbation theory to which we restrict ourselves in this paper. Candidate interactions of six-order perturbation treatments are six-spin interactions involving 2 to 6 sites. One obvious candidate of a six order perturbation treatment is a possible \textit{bicubic} interaction
\begin{equation}
\mathcal{H}_6 \propto \sideset{}{'}\sum_{ij} (\vop{S}_i \cdot \vop{S}_j)^3
\end{equation}
In order to check whether this \textit{bicubic} interaction occurs within higher orders, we have decided to study the system of 2 sites with 3 orbitals up to sixth order in the perturbation. Indeed additional terms occur within the sixth order, which can be explained by the bicubic interaction with a prefactor of $\frac{336 t^6}{(U+2J_\text{H})^5}$. In general, however, it can be assumed that this \textit{bicubic} interaction as well as the other possible six-order terms are small compared to the previously studied second- and fourth-order interactions because it occurs in an even higher order of the perturbation.

\subsection{Spin-models at surfaces due to hopping of electrons beyond nearest neighbor}\label{sec:surfaces}

\begin{figure}
\includegraphics[width=\linewidth]{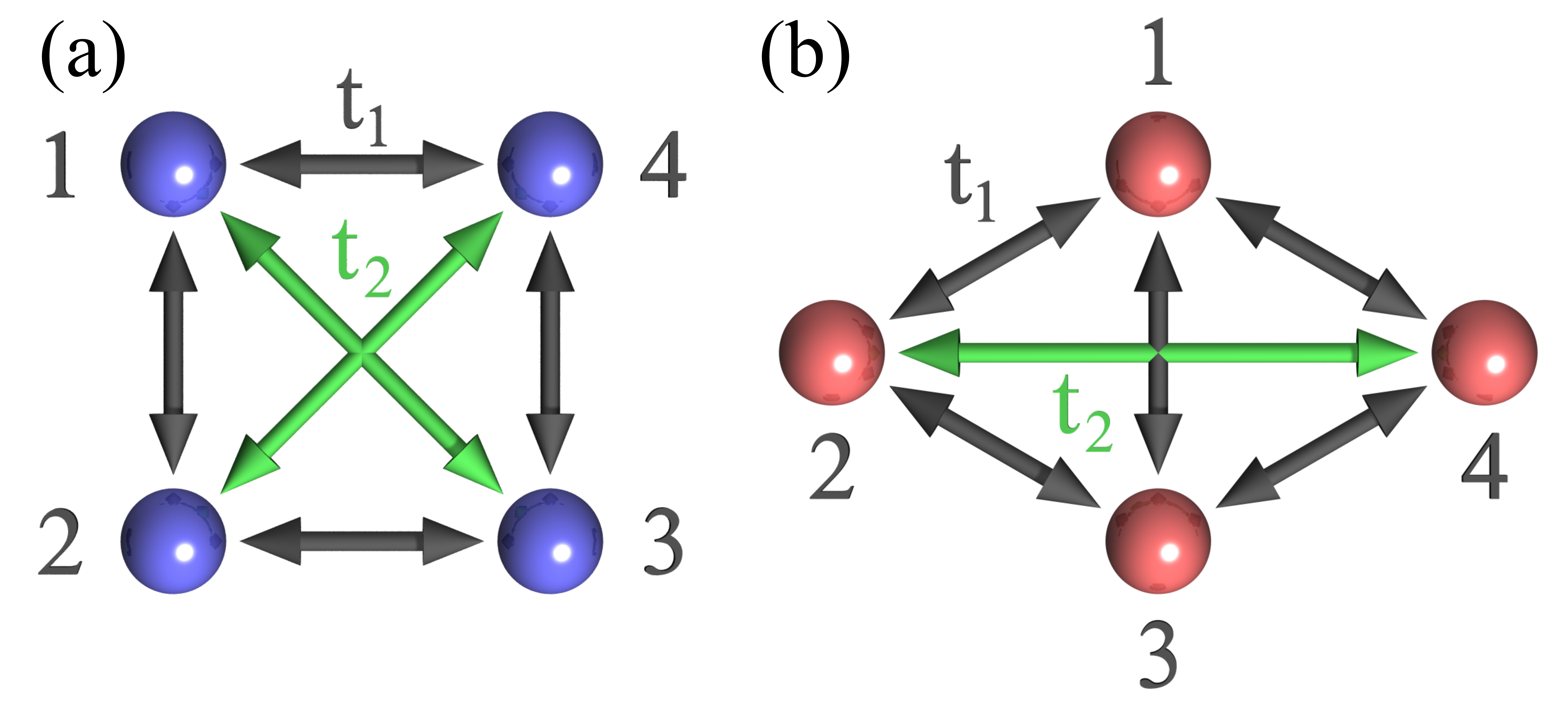}
\caption{Investigated geometries: (a) a square arrangement of atoms as it occurs  \textit{e.g.}\ at the (001)-surface of a bcc or fcc crystal and (b) a hexagonal arrangement as it occurs at the (111)-surface of an fcc crystal or the (0001)-surface of an hexagonal lattice. Sketched are the positions of the atoms and two different neighbor distances. $t_1$ ($t_2$) represents the hopping between nearest (next-nearest) neighbors.}
\label{fig:geometries}
\end{figure}

Up to now, we presented the results assuming that the hopping properties of the electrons between all the atoms are the same. In a real system this is not the case as the value of the hopping parameter $t$ depends mainly on the distance between the two involved atoms, but also on the type of orbitals, the symmetry, the geometry or the environment. In the case of model Hamiltonians describing strongly localized electron systems, the nearest neighbor (NN) approximation  is often sufficient (all hopping parameters $t=0$ between all atom pairs except NN-pairs) and the spin models derived above can be applied practically directly. On the other hand, assuming, the same hopping parameter $t$  is used between all atom pairs, for  example, the interaction between four atoms corresponds to the description of the interactions on a regular tetrahedron. In general, there is a lot of interest in  film, interface or surface geometries of periodic lattices with atom coordinations for which the NN or constant-pair approximation is unrealistic. We want to take this into account and evaluate above spin models for two common types of surfaces, the (001) or (111) oriented surface of fcc crystals using a model with nearest and next-nearest neighbor (NNN) hopping. We focus on a periodic $S=1/2$ system of one atom type with electron interactions involving maximal four lattice sites as indicated in Fig.~\ref{fig:geometries}, where the (001) and (111) geometry are sketched. Obviously, the former represents a square arrangement of the surface atoms, the latter is a triangular or diamond arrangement of an hexagonal lattice.\par

Within both geometries there are two different distances between atoms, the NN- and NNN-distance. Thus, the respective electron hopping is described by two distinct hopping constants, $t_1$ and $t_2$, summarized to $t_\Delta$, with $\Delta\in {1,2}$. 
Going up to second-order  perturbation we find as expected the Heisenberg exchange which reads independent of surface geometry
\begin{equation*}
H_1=-\sideset{}{'}\sum_{i,\delta_\Delta} J_\Delta\, \vop{S}_i\cdot\vop{S}_{i+\delta_\Delta}\qquad\text{with}\qquad J_\Delta = - \frac{2\, t_\Delta^2}{U}\, ,
\end{equation*}
with the respective prefactors $J_1$ ($J_2$) being the exchange constant between NN- (NNN-) pairs. $\sum_{\delta_\Delta}$ denotes the summation over the NN- ($\Delta=1$) and NNN-pairs ($\Delta=2$). $\delta$ denotes the number of NN- (NNN-) pairs. While there are two NNN-pairs for the square lattice, there is only one pair on the hexagonal one, as one of the diagonals (see diagonal connecting atom 1 and 3 in Fig.~\ref{fig:geometries}) is also a NN-pair.

While the second-order expression hold independent on the surface geometry, this is different for fourth-order corrections to the Heisenberg exchange and for the higher-order interactions, where we have to differentiate between the square and the hexagonal lattice.

\subsubsection{(100) surface}
At fourth-order perturbation we expect additional terms to the Heisenberg model proportional to $t^4$. 
For the square lattice,  fourth-order perturbation results  in a correction of $J_1$ by $10t_1^4/U^3$ and a correction of $J_2$ by $(8t_2^4+4 t_1^2 t_2^2-2t_1^4)/U^3$. 
Surprisingly, the correction to $J_2$ does not only contain the naively expected correction proportional to $t_2^4$, but includes also correction terms involving  NN-hopping proportional to $t_1^2\,t_2^2$ and $t_1^4$.
This has its origin  in \eqref{eq:4-sites} where contributions to a ring-hopping involving sites $i$ and $j$ have contributions to the pair-exchange between the spins at sites $i$ and $j$.

For a $S=1/2$ system and fourth-order perturbation in electron hopping we also obtain contributions to the 4-spin interaction, here expressed on a cluster of four sites:
\begin{align}
H_4 = &-K_{1} [\phantom{+}(\vop{S}_1 \cdot \vop{S}_2)\cdot (\vop{S}_3 \cdot \vop{S}_4)\nonumber\\&\hspace{0.84cm}+(\vop{S}_1 \cdot \vop{S}_4)\cdot (\vop{S}_2 \cdot \vop{S}_3)\nonumber\\&\hspace{0.84cm}-(\vop{S}_1 \cdot \vop{S}_3)\cdot (\vop{S}_2 \cdot \vop{S}_4)\ ]\nonumber\\
&-K_{2} \hspace{0.1cm} \hspace{0.25cm}(\vop{S}_1 \cdot \vop{S}_3)\cdot (\vop{S}_2 \cdot \vop{S}_4)
\label{eq:square-4th}
\end{align}
The first term follows the functional form given in \eqref{eq:4-sites} and includes all permutations of exclusive NN-interaction. The related prefactor becomes $K_{1} = -80\,t_1^4/U^3$

More precisely, the variation of the hopping amplitudes between different sites result in preferred paths for a ring-hopping. Thus, we expect additional contributions from the NNN-terms to the 4-spin interaction. We have found that these modifications do not affect the previously discussed NN-ring-hopping but add additional permutations of coupling strength $K_2\propto t_1^2\,t_2^2 $ of the involved sites to the Hamiltonian and can be written in terms of the second term in Eq.~\eqref{eq:square-4th}. Taking a  geometrical picture, this corresponds to a bow-tie-shaped loop which contains two NNN-hopping events and thus hopping terms over the diagonals of the square. We therefore call this term \textit{bow-tie 4-spin term} in the following.
The prefactor of this term was determined to be $K_{2} = -160\,t_1^2\,t_2^2/U^3$.
Thus, the ratio between the prefactors for the two mentioned 4-spin terms is
\begin{equation}
\frac{K_\text{2}}{K_\text{1}} = 2 \left(\frac{t_2}{t_1}\right)^2\ \ \ .
\end{equation}
Depending on the ratio of $t_1$ and $t_2$, the diagonal term can be of the same order of magnitude as the conventional 4-spin term or it might even dominate and should therefore not be neglected in applications of the spin-model, 
\eg Monte-Carlo or spin-dynamic simulations.

Transferring our findings to an infinite lattice, the 4-spin interactions can in general be be written as
\begin{align}
H_4 = - \!\!\sum_{\langle ijkl\rangle_\square}\!\!\!\! \big(K_1[ &(\vop{S}_i \cdot \vop{S}_j) (\vop{S}_k \cdot \vop{S}_l)+(\vop{S}_i \cdot \vop{S}_l) (\vop{S}_j \cdot \vop{S}_k)]\nonumber\\
+\left(K_2 - K_1\right) &(\vop{S}_i \cdot \vop{S}_k)\cdot (\vop{S}_j \cdot \vop{S}_l)\big)\ .
\label{eg:4spin-general}
\end{align}
Here, the notation $\langle ijkl\rangle_{\square}$ denotes sums over unique non-crossing quatruplets of sites of closed loops $i\rightarrow j\rightarrow k\rightarrow l\rightarrow i$.\par
In the case of site-independent hopping amplitude, \ie $t_1=t_2$ and thus $2K_1=K_2$, Eq.~\eqref{eg:4spin-general} simplifies to \eqref{eq:3spin-derived} while is corresponds to Eq.~\eqref{eq:4-spin} in case of pure NN-hopping, \ie $t_2=K_2=0$. 

\subsubsection{(111) surface}

The diamond geometry of the hexagonal (111) lattice offers a different ratio between NNN- and NN-bonds compared to the square lattice (see Fig.~\ref{fig:geometries}), which is at the end reflected in different contributions to the spin Hamiltonian.\par
Collecting all Heisenberg-like terms up to fourth-order perturbation results in the following prefactors:
\begin{align}
J_1 &= -2 \frac{t_1^2}{U} + 8 \frac{t_1^4}{U^3} + 4 \frac{t_1^3t_2}{U^3} - 2 \frac{t_1^2t_2^2}{U^3}\\
J_2 &= -2 \frac{t_2^2}{U} + 8 \frac{t_{2}^4}{U^{3}} + 4 \frac{t_1^3t_2}{U^3} - 2 \frac{t_1^4}{U^3}
\end{align}
The contributions to the 4-spin interactions are equivalent to those for the square lattice (Eq.~\eqref{eq:square-4th}) with the exception of the prefactor of the diagonal 4-spin term which changes to $K_2=-160t_1^3t_2/U^3$. Therefore, the ratio
\begin{equation}
\frac{K_2}{K_1} = 2 \left(\frac{t_2}{t_1}\right)
\end{equation}
makes it even more likely that this term is comparable in size compared to the conventional 4-spin term.\par

\subsection{Importance of three-spin interaction in iron based magnetic thin-film systems}
Now we turn to the description of real magnetic atomic monolayer thick transition-metal films. Magnetic beyond-Heisenberg behavior has been theoretically predicted and experimentally observed for several systems~\cite{Kurz3q,Kobler1996,JPSJ.70.3089,doi:10.1063/1.361678, silke,PhysRevB.92.020401,heinze2011}. The materials specific theoretical modelling of magnetic interactions is generally carried out by means of density functional theory (DFT) and can be pursued along two different paths: (i) The parameters, $t$, $U$, $J_\text{H}$, etc., entering the Hubbard model \eqref{eq:Hamiltonian} are determined directly from DFT and expressions derived above are executed to obtain the exchange parameters of the different magnetic interaction terms. Although it is a possible route, the determination of the Coulomb $U$ and the Hund $J_\text{H}$ parameters have some uncertainties due the screening that should be properly included as a result of those electrons not treated in the Hubbard model explicitly, uncertainties that are sometimes too large to determine the exchange parameters of the spin-model to the level that it is predictable.  (ii) The second approach, which we will follow, is to take the classical limit of the spin models above, \ie  work with classical vector spin, $\mathbf{S}$, instead of vector operators, $\vop{S}$, and calculate the total energies for a large spectrum of magnetic states in momentum or real space using DFT. The spin model parameters are then obtained by comparing the total energy landscape calculated by DFT and the spin model.

\subsubsection{Fe/Rh(111)}

Al-Zubi \textit{et al.}~\cite{PSSB:PSSB201147090} systematically investigated  the magnetism of  Fe monolayers on hexagonal surfaces of different 4$d$ transition-metal substrates using DFT. 
They calculated total energies of a large spectrum of magnetic structures. This included both spin-spiral states for wave vectors $\mathbf{q}$ along the high-symmetry lines of the two-dimensional Brillouin zone and so-called multi-q states of particular $\mathbf{q}$-vectors that allow superpositions of spin spirals of symmetry-equivalent $\mathbf{q}$-vectors. The most remarkable finding was  the prediction of a previously unknown up-up-down-down (\textit{uudd}) state as ground state in Fe/Rh(111), recently confirmed by spin-polarized scanning tunneling microscopy (SP-STM) measurements~\cite{kronlein}. The \textit{uudd} state can be interpreted as interference of 2 spin spirals with wave vectors of opposing directions (2Q-state).

In order to understand the origin of this unknown \textit{uudd} state, they mapped the DFT results onto a spin Hamiltonian, which included  the Heisenberg interaction extended by the biquadratic and the four-spin interaction, the two latter within the nearest-neighbor approximation, and determined the exchange parameters.  The choice of the spin Hamiltonian was taken \textit{ad hoc}, but  motivated by previous successes of similar systems~\cite{Kurz3q,silke,heinze2011}. 
However, they made some puzzling observations. While the energy difference of two unrelated \textit{uudd} states (see Fig.2 of Ref.~\onlinecite{PSSB:PSSB201147090}) characterized by two different wave vectors $\mathbf{q}$ should be the same in comparison to the spin-spiral state (1Q-state) with the corresponding q-vector, \ie,
\begin{equation}
E_{2Q} - E_{1Q} = 4 \, (2K - B)\, ,
\end{equation}
not only the absolute value, but also the sign varied for both.\par
Several attempts were made to resolve this discrepancy, but only the extension of the spin Hamiltonian by the three-spin interaction, which we systematically derived in this paper on grounds of the Hubbard model as an ignored interaction being on the same level as the previously applied biquadratic and 4-spin terms, was able to resolve this issue. 
In fact, depending on the sign of the exchange parameter, the 3-spin interaction selects one of the two \textit{uudd} states to become ground state and indeed it was shown that this explains the magnetic ground state of Fe/Rh(111)~\cite{kronlein}.

\subsubsection{Fe/Ru(0001)}
We show here that a monolayer of Fe deposited on Ru(0001) is a further materials system with beyond-Heisenberg behavior and a system which requires the contribution of the  three-spin interaction in addition to the biquadratic and four-spin interaction for a proper description of the magnetic properties by a spin model. In difference to Fe/Rh(111), DFT calculations of Al-Zubi \textit{et al.}~\cite{PSSB:PSSB201147090} revealed the 120$^\circ$ N\'eel-state (from atom to atom the direction of the magnetic moment changes by 120$^\circ$) as the energetically most favorable of all investigated states, and thus the higher-order interactions do not directly determine the ground state, but the DFT calculations show that this system exhibits a similarly puzzling energy spectrum as Fe/Rh(111) and a proper spin model is required for the description of spin-dynamics, spin excitation and the determination of thermodynamic properties.

In the following we determine the exchange parameter $B$, $Y$, and $K$ of the three beyond-Heisenberg interactions, biquadratic, three-spin, and four-spin, respectively, in the NN-approximation analysing the \textit{ab initio} data of Al-Zubi \textit{et al.}~\cite{PSSB:PSSB201147090}. While the single-wavevector spin spiral (1Q-state) is an eigensolution of the classical Heisenberg model for a periodic lattice, beyond-Heisenberg interactions couple modes of different 1Q-states to multi-Q states with different energy and show that this can result in a more accurate description of Fe/Rh(0001). Therefore, we focus in the following on those single-Q vectors in the two-dimensional hexagonal Brillouin zone of $\mathbf{q}$-vectors defined in reciprocal space as $\mathbf{q}=(q_1,q_2)$ in units of the inplane reciprocal lattice vectors $\mathbf{b}_{1(2)}= (2\pi/a)(1/\sqrt{3},\varpm 1)$, where $a$ is the hexagonal in-plane lattice constant, that can form multi-Q states out of symmetry-equivalent 1Q-states. This includes the high-symmetry point  $\overline{\mathrm{M}}=(1/2,1/2)$ representing  the row-wise antiferromagnetic state, that can form a 3Q-state and the two states $(\overline{\mathrm{\Gamma M}})/2=\pm(1/4,1/4)$ and $3/4(\overline{\mathrm{\Gamma K}})=(\pm1/4,\mp1/4)$ on the high-symmetry lines of the Brillouin zone whose superposition of propagating and counterpropagating waves, \eg $(\overline{\mathrm{\Gamma M}})/2$ and $-(\overline{\mathrm{\Gamma M}})/2$, form 2Q- or \textit{uudd}-states, respectively.

Inserting now the spin structure expressed as a spin-spiral wave, $\mathbf{S}_i=S(\cos(\mathbf{q}\mathbf{R}_i), \sin(\mathbf{q}\mathbf{R}_i), 0)$, where $\mathbf{R}_i$ denotes the position vector to site $i$, for wave vector $\mathbf{q}$, or linear combination of those into the respective expressions for the Heisenberg, biquadratic, three- and four-spin interactions we obtain the following expressions  
\begin{align}
E_{3Q} - E_{\overline{M}} &= \frac{16}{3} (2 K + B - Y)= \phantom{ii}4.6\ \mathrm{meV} \label{eq:diff1}\\
E_{2Q,\frac{\overline{\mathrm{\Gamma M}}}{2}} - E_{\frac{\overline{\mathrm{\Gamma M}}}{2}} &= 4\, (2 K - B - Y)= - 30.3\ \mathrm{meV} \label{eq:diff2}\\
E_{2Q,\frac{3\overline{\mathrm{\Gamma K}}}{4}} - E_{\frac{3\overline{\mathrm{\Gamma K}}}{4}} &= 4\, (2 K - B + Y)= \phantom{-3}7.5\ \mathrm{meV} \label{eq:diff3}
\end{align} 
which  we compared with the energy differences (in meV) obtained from DFT. As one can see, the previously identical energy differences for the two \textit{uudd} states are now separated by $8 Y$ due to the 3-spin interaction. For the  prefactors of the three interactions we obtain:
\begin{equation}
 B = 4.22~\mathrm{meV},\quad\hspace{-1ex}
 Y = 4.73~\mathrm{meV},\quad\hspace{-1ex}
 K = 0.68~\mathrm{meV}   
\end{equation}
The value of the three-spin exchange parameter, $Y$, is in the same order of magnitude as the biquadratic interaction, but is also significantly large compared to the NN-Heisenberg exchange constant $J_1$ ($J_1 = -6.4$ meV)~\cite{PhysRevB.79.094411} and should therefore not be neglected.

Based on our investigation we would argue that the previously puzzling results for Fe/Ru(0001) are the result of the interplay between the biquadratic and a strong 3-spin interaction, which favors one of the magnetic \textit{uudd} textures over the other, an energy difference that could not be resolved before when the three-spin interaction had been neglected.

A final comment on the evaluation of the three-spin interaction. Analogously to the discussion of \eqref{eq:square-4th} and \eqref{eg:4spin-general} the expression \eqref{eq:3spin-derived} can be simplified to
\begin{align}
H_3=- 2Y\sum_{\braket{ijk}_\Delta}  [&(\mathbf{S}_j \cdot \mathbf{S}_i)(\mathbf{S}_i \cdot \mathbf{S}_k)
+(\mathbf{S}_i \cdot \mathbf{S}_j)(\mathbf{S}_j \cdot \mathbf{S}_k)\nonumber\\
+&(\mathbf{S}_i \cdot \mathbf{S}_k)(\mathbf{S}_k \cdot \mathbf{S}_j)]
\end{align}
summing over triangles of NN-sites.

\section{Summary and Conclusions}
In this paper we derived consistently and systematically  the spin Hamiltonian due to interacting electrons up to fourth order perturbation theory in the L\"owdin partitioning algorithm. Starting point was the rotationally invariant multi-orbital Hubbard model that described the interacting electrons on a lattice.  We showed that L\"owdin's  downfolding technique is an efficient approach to map  the effect of the interacting electrons onto an effective spin model. As a result we obtain the spin Hamiltonian
\begin{equation}
H = (H_1 + H_4) \{\text{for}\, S\ge1/2\} + (H_2 + H_3)\{\text{for}\, S\ge 1\}\, ,
\end{equation}
which consists of the Heisenberg Hamiltonian $H_1$ \eqref{eq:Heisenberg}, the biquadratic (four-spin-two-site) $H_2$ \eqref{eq:biquadratic}, the three-spin (four-spin-three-site) $H_3$ \eqref{eq:3-spin}, and the four-spin Hamiltonian (four-spin-four-site) $H_4$ \eqref{eq:4-spin}. The Heisenberg term emerges already in second order perturbation, but the fourth-order perturbation term adds to the exchange coupling parameter. Characteristic of the fourth order terms is the hopping of electrons between 4 orbitals that connect maximally four sites. This form remains correct also for higher spins $S$ treated up the fourth order perturbation theory. On the other hand $S=3/2$ has also 6-order contributions and $S=2$, would have 6- and 8-order contributions, which we have not calculated. Since the dimension of the matrices $H_0$ and $H_1$ in the L\"owdin algorithm grows binomially with the number of orbitals as $\binom{n}{n/2}^2$, the algorithm becomes quickly involved and at the same time the exchange coupling parameters are becoming increasingly smaller and  the terms less important. The exchange coupling parameters of the different Hamiltonians $H_i$, with $i=1,\dots, 4$,  are summarized in detail in the Appendix. 

The spin-orbit interaction was neglected. Subject to the spin-orbit interaction, $S_z$ does not commute anymore with the Hamiltonian, thus the Hamiltonian does not block-diagonalizes anymore for different $m$, and the L\"owdin partitioning becomes more involved.

We showed that our technique is capable of verifying the commonly applied  Heisenberg model, as well as the four-spin and  biquadratic interaction, but unraveled in addition  the occurrence of the three-spin interaction. The importance of the three-spin interaction was verified for the systems of one monolayer Fe on Rh(111) and Ru(0001),  where \textit{ab initio} calculations~\cite{PSSB:PSSB201147090} predicted puzzling  results on the magnetic states that now could be consistently explained. The unusual up-up-down-down ground state stabilized by three-spin interaction in Fe/Rh(111) could recently be confirmed experimentally~\cite{kronlein}.

\section*{Acknowledgments}
We would like to thank Nicholas Ohs for fruitful discussions regarding the development of a versatile L\"owdin partitioning algorithm. We acknowledge funding from the DARPA TEE program through grant MIPR\# HR0011831554 from DOI.

\appendix*
\section{Prefactors for the complete model}
In the main text we focused on presenting the prefactors, or exchange parameter, respectively, of the different spin-models for the simplified case of orbital independent hopping interactions ($t_{i,\alpha,j,\alpha} = t'_{i,\alpha,j,\alpha'} = t$) and for the limits $J' = 0$ and $U' = 0$. Here, we show the  extension of the results for which the hopping interaction between the same  ($t_{i,\alpha,j,\alpha} = t$) and between different orbitals ($t'_{i,\alpha,j,\alpha'} = t'$) are distinct.  Analogously the distinction between intra- and inter-orbital onsite Coulomb repulsion $U$ , and $U'$, respectively, and exchange interaction,  $J$ and $J'$, respectively, is taken into account. Otherwise, all interaction parameters are kept orbital independent for simplicity and remain site independent assuming a periodic lattice of one atom type.\par
In the following, we will denote  exchange parameters as $X_{s\times o}$ with X $\in$ ($J$, $B$, $K$, $Y$) and $s$ and $o$ denoting the number of sites and orbitals, respectively. The prefactors are calculated up to fourth order in the L\"owdin partitioning.
\begin{widetext}
\begin{align*}
J_{2\times 2} = & - \frac{t^2 + t'^2}{U+J_\text{H}} + \frac{4(t^2 + t'^2)^2}{(U+J_\text{H})^3} - \frac{16 t^2 t'^2}{(U+J_\text{H}-U'-J'_\text{H})(U+J_\text{H})^2}  \\
B_{2\times 2} = & - \frac{2(t^2+t'^2)^2}{(U+J_\text{H})^3} + \frac{(t^2 - t'^2)^2}{2(U+J_\text{H})^2 J_\text{H}} + \frac{4(t^2 - t'^2)^2}{(2U+U')(U+J_\text{H})^2} + \frac{t^4 - 14 t^2 t'^2 + t'^2}{(U+J_\text{H}-U'-J'_\text{H})(U+J_\text{H})^2} \\ & + \frac{(t^2 - t'^2)^2}{(U+J_\text{H}-U'+J'_\text{H})(U+J_\text{H})^2} \\
\ \\
\hline
\ \\
J_{3\times 2} = & - \frac{t^2 + t'^2}{U+J_\text{H}} +\frac{12(t^2+t'^2)^2}{(U+J_\text{H})^3}-\frac{3(t^4+6t^2t'^2+t'^4)}{(2U+2J_\text{H}-U'-J'_\text{H})(U+J_\text{H})^2}\\
& - (t^2-t'^2)^2 \cdot \bigg(-\frac{27}{4J_\text{H}^2(U+J_\text{H})}+\frac{12}{J_\text{H}^2 (2U+J_\text{H})}+\frac{3}{4 J_\text{H}^2 (U+3 J_\text{H})} -\frac{3}{2 J_\text{H} (U+J_\text{H})^2}\\
& \hspace*{2.44cm}+\frac{3}{(2U+2J_\text{H}-U'+J') (U+J_\text{H})^2} +\frac{3}{(2U+J_\text{H}+U')(U+J_\text{H})^2}\bigg)\\
\ \\
B_{3\times 2} = & -(t^2+t'^2)^2 \cdot \Big(+\frac{2}{(U+J_\text{H})^3}+\frac{3}{(U+J_\text{H}-U'-J'_\text{H})(U+J_\text{H})^2}\Big)\\
& -(t^2-t'^2)^2 \cdot \bigg(-\frac{1}{2J_\text{H}(U+J_\text{H})^2}-\frac{1}{(U+J_\text{H}-U'+J'_\text{H})(U+J_\text{H})^2}\\
& \hspace*{2.44cm}-\frac{4}{(2U+U')(U+J_\text{H})^2}-\frac{4}{(U+J_\text{H}-U'-J'_\text{H})(U+J_\text{H})^2}\bigg)\\
\ \\
Y_{3\times 2} = & -\frac{16t^2t'^2}{(U+J_\text{H}-U'-J'_\text{H})(U+J_\text{H})^2}+\frac{2t^4+12t^2t'^2+2t'^4}{(2U+2J_\text{H}-U'-J'_\text{H})(U+J_\text{H})^2} -\frac{6t^4+20t^2t'^2+6t'^4}{(U+J_\text{H})^3}\\
& -(t^2-t'^2)^2\cdot \bigg(-\frac{1}{2J_\text{H}^2(U+3J_\text{H})}+\frac{1}{J_\text{H}(U+J_\text{H})^2}+\frac{9}{2J_\text{H}^2(U+J_\text{H})}-\frac{8}{J_\text{H}^2(2U+J_\text{H})}\\
& \hspace*{2.44cm} -\frac{2}{(2U+J_\text{H}+U')(U+J_\text{H})^2}-\frac{2}{(2U+2J_\text{H}-U'+J'_\text{H})(U+J_\text{H})^2}\bigg)\\
\ \\
\hline
\ \\
J_{4\times 2} = & - \frac{t^2 + t'^2}{U+J_\text{H}}\\
& -(t^2-t'^2)^2\cdot \bigg(+\frac{24}{J_\text{H}^2(2U+J_\text{H})}-\frac{27}{2J_\text{H}^2(U+J_\text{H})}+\frac{3}{2J_\text{H}^2(3J_\text{H}+U)}-\frac{3}{J_\text{H}(U+J_\text{H})^2}\\
& \hspace*{2.45cm}+\frac{6}{(2U+2J_\text{H}-U'+J'_\text{H})(U+J_\text{H})^2}+\frac{6}{(2U+J_\text{H}+U')(U+J_\text{H})^2}\bigg)\\
& +\frac{23t^4+58t^2t'^2+23t'^4}{(U+J_\text{H})^3}-\frac{6t^4+36t^2t'^2+6t'^4}{(2U+2J_\text{H}-U'-J'_\text{H})(U+J_\text{H})^2}+\frac{16t^2t'^2}{(U+J_\text{H}-U'-J'_\text{H})(U+J_\text{H})^2}\\
\ \\
B_{4\times 2} = & -(t^2-t'^2)^2\cdot \bigg(-\frac{1}{2J_\text{H}(U+J_\text{H})^2}-\frac{4}{(2U+U')(U+J_\text{H})^2}-\frac{1}{(U+J_\text{H}-U'+J'_\text{H})(U+J_\text{H})^2}\bigg)\\
&\hspace*{2.45cm}-\frac{2(t^4+2t^2t'^2+t'^4)}{(U+J_\text{H})^3}+\frac{t^4-14t^2t'^2+t'^4}{(U+J_\text{H}-U'-J'_\text{H})(U+J_\text{H})^2}\\
\ \\
Y_{4\times 2} = & -(t^2-t'^2)^2\cdot \bigg(-\frac{2}{(2U+2J_\text{H}-U'+J'_\text{H})(U+J_\text{H})^2}-\frac{2}{(2U+J_\text{H}+U')(U+J_\text{H})^2}-\frac{8}{(2U+J_\text{H})J_\text{H}^2}\\
&\hspace*{2.5cm}+\frac{9}{2(U+J_\text{H})J_\text{H}^2}-\frac{1}{2J_\text{H}^2(U+3J_\text{H})}+\frac{1}{(U+J_\text{H})^2J_\text{H}}\bigg)\\
&+\frac{2t^4+12t^2t'^2+2t'^4}{(2U+2J_\text{H}-U'-J'_\text{H})(U+J_\text{H})^2}-\frac{16t^2t'^2}{(U+J_\text{H}-U'-J'_\text{H})(U+J_\text{H})^2}-\frac{6t^4+20t^2t'^2+6t'^4}{(U+J_\text{H})^3}\\
\ \\
K_{4\times 2}= & -\frac{5(t^4+6t^2(t'^2)+t'^4)}{4(U+J_\text{H})^3}\\
\ \\
\hline
\ \\
J_{2\times 3} = & -\frac{2}{3}\frac{(t^2+2t'^2)}{(U+2J_\text{H})} -\frac{8t'^2(t-t')^2}{3(U+5J_\text{H}-U'-J'_\text{H})(U+2J_\text{H})^2}-\frac{(t^2+2tt'+3t'^2)(t-t')^2}{(U+J_\text{H}+U')(U+2J_\text{H})^2} +\frac{4(t^2+2t'^2)^2}{(U+2J_\text{H})^3}\\
& - \frac{(t^2+2tt'+3t'^2)(t-t')^2}{2(U+2J_\text{H}-U'+J'_\text{H})(U+2J_\text{H})^2}-\frac{3t^4+76t^2t'^2+76tt'^3+25t'^4}{6(U+2J_\text{H}-U'-J'_\text{H})(U+2J_\text{H})^2}-\frac{(t^2-t'^2)^2}{9(U+2J_\text{H})^2J_\text{H}}\\
\ \\
B_{2\times 3} = & +\frac{(t^2+2tt'+3t'^2)(t-t')^2}{3(U+2J_\text{H}-U'+J'_\text{H})(U+2J_\text{H})^2}+\frac{3t^4-52t^2t'^2-52tt'^3-7t'^4}{9(U+2J_\text{H}-U'-J'_\text{H})(U+2J_\text{H})^2}\\
& +\frac{16t'^2(t-t')^2}{9(U+5J_\text{H}-U'-J'_\text{H})(U+2J_\text{H})^2}+\frac{2(t^2+2tt'+3t'^2)(t-t')^2}{3(U+J_\text{H}+U')(U+2J_\text{H})^2}\\
& +\frac{2(t-t')^2(t+t')^2}{27(U+2J_\text{H})^2J_\text{H}}+\frac{8(t^2+2t'^2)^2}{9(U+2J_\text{H})^3}\\
\ \\
\hline
\ \\
J_{3\times 3} = & -\frac{2}{3}\frac{(t^2+2t'^2)}{(U+2J_\text{H})}-\frac{-3 t^4+20 t^2 t'^2+20 t t'^3-t'^4}{6 (U+2 J_\text{H}-U'-J'_\text{H}) (U+2 J_\text{H})^2}-\frac{-110 t^4-456 t^2 t'^2-16 t t'^3-444 t'^4}{9 (U+2 J_\text{H})^3}\\
& -\frac{-10 t^4-80 t^2 t'^2-40 t t'^3-50 t'^4}{3 (2 U+4 J_\text{H}-U'-J'_\text{H}) (U+2 J_\text{H})^2}\\
&-(t-t')^2\cdot \bigg[+\frac{t^2+2tt'+3t'^2}{(U+2J_\text{H})^2} \cdot\bigg( -\frac{20(U+2J_\text{H})}{27 J_\text{H} (5J_\text{H}+U)}+\frac{1}{U+J_\text{H}+U'}+\frac{20}{3(2U+3J_\text{H}+U')}\\
&\hspace*{5.56cm}+\frac{1}{2(U+2J_\text{H}-U'+J'_\text{H})}+\frac{10}{3(2U+4J_\text{H}-U'+J'_\text{H})}\bigg)\\
&\hspace*{2.09cm}+\frac{23 t^2+46 t t'+63 t'^2}{27 (U+2 J_\text{H})^2 J_\text{H}}-\frac{8  t'^2}{3 (U+5 J_\text{H}-U'-J'_\text{H}) (U+2 J_\text{H})^2}\bigg]\\
\ \\
B_{3\times 3} = &+\frac{(t-t')^2}{(U+2J_\text{H})^2} \cdot \bigg(\frac{2t^2+4tt'+6t'^2}{3(U+J_\text{H}+U')}+\frac{16t'^2}{9(U+5J_\text{H}-U'-J'_\text{H})}+\frac{t^2+2tt'+3t'^2}{3(U+2J_\text{H}-U'+J'_\text{H})}+\frac{2(t+t')^2}{27 J_\text{H}}\bigg)\\
&+\frac{3t^4-52t^2t'^2-52tt'^3-7t'^4}{9(U+2J_\text{H}-U'-J'_\text{H})(U+2J_\text{H})^2}+\frac{8(t^2+2t'^2)^2}{9(U+2J_\text{H})^3}\\
Y_{3\times 3} = &-(t-t')^2\cdot(t^2+2tt'+3t'^2)\cdot \bigg[-\frac{1}{(U+2J_\text{H})^2}\cdot \bigg(\frac{16}{9(2U+3J_\text{H}+U')}+\frac{8}{9(2U+4J_\text{H}-U'+J'_\text{H})}+\frac{16}{81J_\text{H}}\bigg)\\
&\hspace*{4.8cm}+\frac{16}{243(U+2J_\text{H})J_\text{H}^2}-\frac{16}{243(5J_\text{H}+U)J_\text{H}^2}\bigg]\\
&-\frac{1}{(U+2J_\text{H})^2}\cdot \bigg[-\frac{8t^4+64t^2t'^2+32tt'^3+40t'^4}{9(2U+4J_\text{H}-U'-J'_\text{H})}+\frac{16t'^2(t'+2t)^2}{9(U+2J_\text{H}-U'-J'_\text{H})}\\
&\hspace*{2.75cm}+\frac{64t^4+384t^2t'^2+128tt'^3+288t'^4}{27(U+2J_\text{H})}\bigg]
 \end{align*}
\end{widetext}


\begin{thebibliography}{63}%
\makeatletter
\providecommand \@ifxundefined [1]{%
 \@ifx{#1\undefined}
}%
\providecommand \@ifnum [1]{%
 \ifnum #1\expandafter \@firstoftwo
 \else \expandafter \@secondoftwo
 \fi
}%
\providecommand \@ifx [1]{%
 \ifx #1\expandafter \@firstoftwo
 \else \expandafter \@secondoftwo
 \fi
}%
\providecommand \natexlab [1]{#1}%
\providecommand \enquote  [1]{``#1''}%
\providecommand \bibnamefont  [1]{#1}%
\providecommand \bibfnamefont [1]{#1}%
\providecommand \citenamefont [1]{#1}%
\providecommand \href@noop [0]{\@secondoftwo}%
\providecommand \href [0]{\begingroup \@sanitize@url \@href}%
\providecommand \@href[1]{\@@startlink{#1}\@@href}%
\providecommand \@@href[1]{\endgroup#1\@@endlink}%
\providecommand \@sanitize@url [0]{\catcode `\\12\catcode `\$12\catcode
  `\&12\catcode `\#12\catcode `\^12\catcode `\_12\catcode `\%12\relax}%
\providecommand \@@startlink[1]{}%
\providecommand \@@endlink[0]{}%
\providecommand \url  [0]{\begingroup\@sanitize@url \@url }%
\providecommand \@url [1]{\endgroup\@href {#1}{\urlprefix }}%
\providecommand \urlprefix  [0]{URL }%
\providecommand \Eprint [0]{\href }%
\providecommand \doibase [0]{https://doi.org/}%
\providecommand \selectlanguage [0]{\@gobble}%
\providecommand \bibinfo  [0]{\@secondoftwo}%
\providecommand \bibfield  [0]{\@secondoftwo}%
\providecommand \translation [1]{[#1]}%
\providecommand \BibitemOpen [0]{}%
\providecommand \bibitemStop [0]{}%
\providecommand \bibitemNoStop [0]{.\EOS\space}%
\providecommand \EOS [0]{\spacefactor3000\relax}%
\providecommand \BibitemShut  [1]{\csname bibitem#1\endcsname}%
\let\auto@bib@innerbib\@empty
\bibitem [{\citenamefont {Anderson}(1963)}]{Anderson:1963}%
  \BibitemOpen
  \bibfield  {author} {\bibinfo {author} {\bibfnamefont {P.~W.}\ \bibnamefont
  {Anderson}},\ }\href@noop {} {\emph {\bibinfo {title} {Exchange in
  Insulators: Superexchange, Direct Exchange, and Double Exchange}}},\ edited
  by\ \bibinfo {editor} {\bibfnamefont {G.}~\bibnamefont {Rado}}\ and\ \bibinfo
  {editor} {\bibfnamefont {H.}~\bibnamefont {Suhl}},\ Vol.~\bibinfo {volume}
  {1}\ (\bibinfo  {publisher} {Magnetism Academic},\ \bibinfo {year}
  {1963})\BibitemShut {NoStop}%
\bibitem [{\citenamefont {Zeiger}\ and\ \citenamefont
  {Pratt}(1973)}]{Zeiger:1973}%
  \BibitemOpen
  \bibfield  {author} {\bibinfo {author} {\bibfnamefont {H.~J.}\ \bibnamefont
  {Zeiger}}\ and\ \bibinfo {author} {\bibfnamefont {G.~W.}\ \bibnamefont
  {Pratt}},\ }\href@noop {} {\emph {\bibinfo {title} {Magnetic Interactions in
  Solids}}},\ Monographs on Physics\ (\bibinfo  {publisher} {Oxford University
  Press},\ \bibinfo {year} {1973})\BibitemShut {NoStop}%
\bibitem [{\citenamefont {de~Graaf}\ and\ \citenamefont
  {Broer}(2015)}]{Graaf:2015}%
  \BibitemOpen
  \bibfield  {author} {\bibinfo {author} {\bibfnamefont {C.}~\bibnamefont
  {de~Graaf}}\ and\ \bibinfo {author} {\bibfnamefont {R.}~\bibnamefont
  {Broer}},\ }\href@noop {} {\emph {\bibinfo {title} {Magnetic Interactions in
  Solids and Molecules}}},\ Theoretical Chemistry and Computational Modelling\
  (\bibinfo  {publisher} {Springer International Publishing},\ \bibinfo {year}
  {2015})\BibitemShut {NoStop}%
\bibitem [{\citenamefont {Fert}\ \emph {et~al.}(2017)\citenamefont {Fert},
  \citenamefont {Reyen},\ and\ \citenamefont {Cros}}]{Fert:17}%
  \BibitemOpen
  \bibfield  {author} {\bibinfo {author} {\bibfnamefont {A.}~\bibnamefont
  {Fert}}, \bibinfo {author} {\bibfnamefont {N.}~\bibnamefont {Reyen}},\ and\
  \bibinfo {author} {\bibfnamefont {V.}~\bibnamefont {Cros}},\ }\href@noop {}
  {\bibfield  {journal} {\bibinfo  {journal} {Nat. Rev. Mat.}\ }\textbf
  {\bibinfo {volume} {2}},\ \bibinfo {pages} {17031} (\bibinfo {year}
  {2017})}\BibitemShut {NoStop}%
\bibitem [{\citenamefont {Shindou}\ and\ \citenamefont
  {Nagaosa}(2001)}]{Nagaosa:01}%
  \BibitemOpen
  \bibfield  {author} {\bibinfo {author} {\bibfnamefont {R.}~\bibnamefont
  {Shindou}}\ and\ \bibinfo {author} {\bibfnamefont {N.}~\bibnamefont
  {Nagaosa}},\ }\href@noop {} {\bibfield  {journal} {\bibinfo  {journal} {Phys.
  Rev. Lett.}\ }\textbf {\bibinfo {volume} {87}},\ \bibinfo {pages} {116801}
  (\bibinfo {year} {2001})}\BibitemShut {NoStop}%
\bibitem [{\citenamefont {Schulz}\ \emph {et~al.}(2012)\citenamefont {Schulz},
  \citenamefont {Ritz}, \citenamefont {Bauer}, \citenamefont {Halder},
  \citenamefont {Wagner}, \citenamefont {Franz}, \citenamefont {Pfleiderer},
  \citenamefont {Everschor}, \citenamefont {M.},\ and\ \citenamefont
  {A.}}]{Schultz:12}%
  \BibitemOpen
  \bibfield  {author} {\bibinfo {author} {\bibfnamefont {T.}~\bibnamefont
  {Schulz}}, \bibinfo {author} {\bibfnamefont {R.}~\bibnamefont {Ritz}},
  \bibinfo {author} {\bibfnamefont {A.}~\bibnamefont {Bauer}}, \bibinfo
  {author} {\bibfnamefont {M.}~\bibnamefont {Halder}}, \bibinfo {author}
  {\bibfnamefont {M.}~\bibnamefont {Wagner}}, \bibinfo {author} {\bibfnamefont
  {C.}~\bibnamefont {Franz}}, \bibinfo {author} {\bibfnamefont
  {C.}~\bibnamefont {Pfleiderer}}, \bibinfo {author} {\bibfnamefont
  {K.}~\bibnamefont {Everschor}}, \bibinfo {author} {\bibfnamefont
  {G.}~\bibnamefont {M.}},\ and\ \bibinfo {author} {\bibfnamefont
  {R.}~\bibnamefont {A.}},\ }\href@noop {} {\bibfield  {journal} {\bibinfo
  {journal} {Nat. Phys.}\ }\textbf {\bibinfo {volume} {8}},\ \bibinfo {pages}
  {301} (\bibinfo {year} {2012})}\BibitemShut {NoStop}%
\bibitem [{\citenamefont {dos Santos~Dias}\ \emph {et~al.}(2016)\citenamefont
  {dos Santos~Dias}, \citenamefont {Bouaziz}, \citenamefont {Bouhassoune},
  \citenamefont {Bl\"ugel},\ and\ \citenamefont {Samir}}]{Diea:17}%
  \BibitemOpen
  \bibfield  {author} {\bibinfo {author} {\bibfnamefont {M.}~\bibnamefont {dos
  Santos~Dias}}, \bibinfo {author} {\bibfnamefont {J.}~\bibnamefont {Bouaziz}},
  \bibinfo {author} {\bibfnamefont {M.}~\bibnamefont {Bouhassoune}}, \bibinfo
  {author} {\bibfnamefont {S.}~\bibnamefont {Bl\"ugel}},\ and\ \bibinfo
  {author} {\bibfnamefont {L.}~\bibnamefont {Samir}},\ }\href@noop {}
  {\bibfield  {journal} {\bibinfo  {journal} {Nat. Commun.}\ }\textbf {\bibinfo
  {volume} {7}},\ \bibinfo {pages} {13613} (\bibinfo {year}
  {2016})}\BibitemShut {NoStop}%
\bibitem [{\citenamefont {Dup{\'e}}\ \emph {et~al.}(2014)\citenamefont
  {Dup{\'e}}, \citenamefont {Hoffmann}, \citenamefont {Paillard},\ and\
  \citenamefont {Heinze}}]{ncomms5030}%
  \BibitemOpen
  \bibfield  {author} {\bibinfo {author} {\bibfnamefont {B.}~\bibnamefont
  {Dup{\'e}}}, \bibinfo {author} {\bibfnamefont {M.}~\bibnamefont {Hoffmann}},
  \bibinfo {author} {\bibfnamefont {C.}~\bibnamefont {Paillard}},\ and\
  \bibinfo {author} {\bibfnamefont {S.}~\bibnamefont {Heinze}},\ }\href@noop {}
  {\bibfield  {journal} {\bibinfo  {journal} {Nat. Commun.}\ }\textbf {\bibinfo
  {volume} {5}},\ \bibinfo {pages} {4030} (\bibinfo {year} {2014})}\BibitemShut
  {NoStop}%
\bibitem [{\citenamefont {De\'ak}\ \emph {et~al.}(2011)\citenamefont {De\'ak},
  \citenamefont {Szunyogh},\ and\ \citenamefont
  {Ujfalussy}}]{PhysRevB.84.224413}%
  \BibitemOpen
  \bibfield  {author} {\bibinfo {author} {\bibfnamefont {A.}~\bibnamefont
  {De\'ak}}, \bibinfo {author} {\bibfnamefont {L.}~\bibnamefont {Szunyogh}},\
  and\ \bibinfo {author} {\bibfnamefont {B.}~\bibnamefont {Ujfalussy}},\
  }\href@noop {} {\bibfield  {journal} {\bibinfo  {journal} {Phys. Rev. B}\
  }\textbf {\bibinfo {volume} {84}},\ \bibinfo {pages} {224413} (\bibinfo
  {year} {2011})}\BibitemShut {NoStop}%
\bibitem [{\citenamefont {Hoffmann}\ \emph {et~al.}(2015)\citenamefont
  {Hoffmann}, \citenamefont {Weischenberg}, \citenamefont {Dup\'e},
  \citenamefont {Freimuth}, \citenamefont {Ferriani}, \citenamefont
  {Mokrousov},\ and\ \citenamefont {Heinze}}]{PhysRevB.92.020401}%
  \BibitemOpen
  \bibfield  {author} {\bibinfo {author} {\bibfnamefont {M.}~\bibnamefont
  {Hoffmann}}, \bibinfo {author} {\bibfnamefont {J.}~\bibnamefont
  {Weischenberg}}, \bibinfo {author} {\bibfnamefont {B.}~\bibnamefont
  {Dup\'e}}, \bibinfo {author} {\bibfnamefont {F.}~\bibnamefont {Freimuth}},
  \bibinfo {author} {\bibfnamefont {P.}~\bibnamefont {Ferriani}}, \bibinfo
  {author} {\bibfnamefont {Y.}~\bibnamefont {Mokrousov}},\ and\ \bibinfo
  {author} {\bibfnamefont {S.}~\bibnamefont {Heinze}},\ }\href@noop {}
  {\bibfield  {journal} {\bibinfo  {journal} {Phys. Rev. B}\ }\textbf {\bibinfo
  {volume} {92}},\ \bibinfo {pages} {020401(R)} (\bibinfo {year}
  {2015})}\BibitemShut {NoStop}%
\bibitem [{\citenamefont {Gyorffy}\ \emph {et~al.}(1985)\citenamefont
  {Gyorffy}, \citenamefont {Pindor}, \citenamefont {Staunton}, \citenamefont
  {Stocks},\ and\ \citenamefont {Winter}}]{gyorffy1985}%
  \BibitemOpen
  \bibfield  {author} {\bibinfo {author} {\bibfnamefont {B.}~\bibnamefont
  {Gyorffy}}, \bibinfo {author} {\bibfnamefont {A.}~\bibnamefont {Pindor}},
  \bibinfo {author} {\bibfnamefont {J.}~\bibnamefont {Staunton}}, \bibinfo
  {author} {\bibfnamefont {G.}~\bibnamefont {Stocks}},\ and\ \bibinfo {author}
  {\bibfnamefont {H.}~\bibnamefont {Winter}},\ }\href@noop {} {\bibfield
  {journal} {\bibinfo  {journal} {J. Phys. F}\ }\textbf {\bibinfo {volume}
  {15}},\ \bibinfo {pages} {1337} (\bibinfo {year} {1985})}\BibitemShut
  {NoStop}%
\bibitem [{\citenamefont {Halilov}\ \emph {et~al.}(1998)\citenamefont
  {Halilov}, \citenamefont {Eschrig}, \citenamefont {Perlov},\ and\
  \citenamefont {Oppeneer}}]{PhysRevB.58.293}%
  \BibitemOpen
  \bibfield  {author} {\bibinfo {author} {\bibfnamefont {S.~V.}\ \bibnamefont
  {Halilov}}, \bibinfo {author} {\bibfnamefont {H.}~\bibnamefont {Eschrig}},
  \bibinfo {author} {\bibfnamefont {A.~Y.}\ \bibnamefont {Perlov}},\ and\
  \bibinfo {author} {\bibfnamefont {P.~M.}\ \bibnamefont {Oppeneer}},\ }\href
  {https://doi.org/10.1103/PhysRevB.58.293} {\bibfield  {journal} {\bibinfo
  {journal} {Phys. Rev. B}\ }\textbf {\bibinfo {volume} {58}},\ \bibinfo
  {pages} {293} (\bibinfo {year} {1998})}\BibitemShut {NoStop}%
\bibitem [{\citenamefont {Pajda}\ \emph {et~al.}(2001)\citenamefont {Pajda},
  \citenamefont {Kudrnovsk\'y}, \citenamefont {Turek}, \citenamefont {Drchal},\
  and\ \citenamefont {Bruno}}]{PhysRevB.64.174402}%
  \BibitemOpen
  \bibfield  {author} {\bibinfo {author} {\bibfnamefont {M.}~\bibnamefont
  {Pajda}}, \bibinfo {author} {\bibfnamefont {J.}~\bibnamefont {Kudrnovsk\'y}},
  \bibinfo {author} {\bibfnamefont {I.}~\bibnamefont {Turek}}, \bibinfo
  {author} {\bibfnamefont {V.}~\bibnamefont {Drchal}},\ and\ \bibinfo {author}
  {\bibfnamefont {P.}~\bibnamefont {Bruno}},\ }\href
  {https://doi.org/10.1103/PhysRevB.64.174402} {\bibfield  {journal} {\bibinfo
  {journal} {Phys. Rev. B}\ }\textbf {\bibinfo {volume} {64}},\ \bibinfo
  {pages} {174402} (\bibinfo {year} {2001})}\BibitemShut {NoStop}%
\bibitem [{\citenamefont {F{\"a}hnle}\ and\ \citenamefont
  {Steiauf}(2007)}]{fahnle2007}%
  \BibitemOpen
  \bibfield  {author} {\bibinfo {author} {\bibfnamefont {M.}~\bibnamefont
  {F{\"a}hnle}}\ and\ \bibinfo {author} {\bibfnamefont {D.}~\bibnamefont
  {Steiauf}},\ }\href@noop {} {\emph {\bibinfo {title} {Dissipative
  magnetization dynamics close to the adiabatic regime}}}\ (\bibinfo
  {publisher} {Wiley Online Library},\ \bibinfo {year} {2007})\BibitemShut
  {NoStop}%
\bibitem [{\citenamefont {Singer}\ \emph
  {et~al.}(2011{\natexlab{a}})\citenamefont {Singer}, \citenamefont
  {Dietermann},\ and\ \citenamefont {F\"ahnle}}]{PhysRevLett.107.017204}%
  \BibitemOpen
  \bibfield  {author} {\bibinfo {author} {\bibfnamefont {R.}~\bibnamefont
  {Singer}}, \bibinfo {author} {\bibfnamefont {F.}~\bibnamefont {Dietermann}},\
  and\ \bibinfo {author} {\bibfnamefont {M.}~\bibnamefont {F\"ahnle}},\ }\href
  {https://doi.org/10.1103/PhysRevLett.107.017204} {\bibfield  {journal}
  {\bibinfo  {journal} {Phys. Rev. Lett.}\ }\textbf {\bibinfo {volume} {107}},\
  \bibinfo {pages} {017204} (\bibinfo {year} {2011}{\natexlab{a}})}\BibitemShut
  {NoStop}%
\bibitem [{\citenamefont {Singer}\ \emph
  {et~al.}(2011{\natexlab{b}})\citenamefont {Singer}, \citenamefont
  {Dietermann},\ and\ \citenamefont {F\"ahnle}}]{PhysRevLett.107.119901}%
  \BibitemOpen
  \bibfield  {author} {\bibinfo {author} {\bibfnamefont {R.}~\bibnamefont
  {Singer}}, \bibinfo {author} {\bibfnamefont {F.}~\bibnamefont {Dietermann}},\
  and\ \bibinfo {author} {\bibfnamefont {M.}~\bibnamefont {F\"ahnle}},\ }\href
  {https://doi.org/10.1103/PhysRevLett.107.119901} {\bibfield  {journal}
  {\bibinfo  {journal} {Phys. Rev. Lett.}\ }\textbf {\bibinfo {volume} {107}},\
  \bibinfo {pages} {119901(E)} (\bibinfo {year}
  {2011}{\natexlab{b}})}\BibitemShut {NoStop}%
\bibitem [{\citenamefont {Heisenberg}(1928)}]{Heisenberg:1928}%
  \BibitemOpen
  \bibfield  {author} {\bibinfo {author} {\bibfnamefont {W.}~\bibnamefont
  {Heisenberg}},\ }\href@noop {} {\bibfield  {journal} {\bibinfo  {journal}
  {Zeitschrift f{\"u}r Physik}\ }\textbf {\bibinfo {volume} {49}},\ \bibinfo
  {pages} {619} (\bibinfo {year} {1928})}\BibitemShut {NoStop}%
\bibitem [{\citenamefont {Bode}\ \emph {et~al.}(2007)\citenamefont {Bode},
  \citenamefont {Heide}, \citenamefont {von Bergmann}, \citenamefont
  {Ferriani}, \citenamefont {Heinze}, \citenamefont {Bihlmayer}, \citenamefont
  {Kubetzka}, \citenamefont {Pietzsch}, \citenamefont {Bl{\"u}gel},\ and\
  \citenamefont {Wiesendanger}}]{nature05802}%
  \BibitemOpen
  \bibfield  {author} {\bibinfo {author} {\bibfnamefont {M.}~\bibnamefont
  {Bode}}, \bibinfo {author} {\bibfnamefont {M.}~\bibnamefont {Heide}},
  \bibinfo {author} {\bibfnamefont {K.}~\bibnamefont {von Bergmann}}, \bibinfo
  {author} {\bibfnamefont {P.}~\bibnamefont {Ferriani}}, \bibinfo {author}
  {\bibfnamefont {S.}~\bibnamefont {Heinze}}, \bibinfo {author} {\bibfnamefont
  {G.}~\bibnamefont {Bihlmayer}}, \bibinfo {author} {\bibfnamefont
  {A.}~\bibnamefont {Kubetzka}}, \bibinfo {author} {\bibfnamefont
  {O.}~\bibnamefont {Pietzsch}}, \bibinfo {author} {\bibfnamefont
  {S.}~\bibnamefont {Bl{\"u}gel}},\ and\ \bibinfo {author} {\bibfnamefont
  {R.}~\bibnamefont {Wiesendanger}},\ }\href@noop {} {\bibfield  {journal}
  {\bibinfo  {journal} {Nature}\ }\textbf {\bibinfo {volume} {447}},\ \bibinfo
  {pages} {190} (\bibinfo {year} {2007})}\BibitemShut {NoStop}%
\bibitem [{\citenamefont {Turek}\ \emph {et~al.}(2006)\citenamefont {Turek},
  \citenamefont {Kudrnovský}, \citenamefont {Drchal},\ and\ \citenamefont
  {Bruno}}]{doi:10.1080/14786430500504048}%
  \BibitemOpen
  \bibfield  {author} {\bibinfo {author} {\bibfnamefont {I.}~\bibnamefont
  {Turek}}, \bibinfo {author} {\bibfnamefont {J.}~\bibnamefont {Kudrnovský}},
  \bibinfo {author} {\bibfnamefont {V.}~\bibnamefont {Drchal}},\ and\ \bibinfo
  {author} {\bibfnamefont {P.}~\bibnamefont {Bruno}},\ }\href@noop {}
  {\bibfield  {journal} {\bibinfo  {journal} {Philos. Mag.}\ }\textbf {\bibinfo
  {volume} {86}},\ \bibinfo {pages} {1713} (\bibinfo {year}
  {2006})}\BibitemShut {NoStop}%
\bibitem [{\citenamefont {\ifmmode \mbox{\c{S}}\else \c{S}\fi{}a\ifmmode
  \mbox{\c{s}}\else \c{s}\fi{}\ifmmode \imath \else \i
  \fi{}o\ifmmode~\breve{g}\else \u{g}\fi{}lu}\ \emph
  {et~al.}(2005)\citenamefont {\ifmmode \mbox{\c{S}}\else \c{S}\fi{}a\ifmmode
  \mbox{\c{s}}\else \c{s}\fi{}\ifmmode \imath \else \i
  \fi{}o\ifmmode~\breve{g}\else \u{g}\fi{}lu}, \citenamefont {Sandratskii},
  \citenamefont {Bruno},\ and\ \citenamefont {Galanakis}}]{PhysRevB.72.184415}%
  \BibitemOpen
  \bibfield  {author} {\bibinfo {author} {\bibfnamefont {E.}~\bibnamefont
  {\ifmmode \mbox{\c{S}}\else \c{S}\fi{}a\ifmmode \mbox{\c{s}}\else
  \c{s}\fi{}\ifmmode \imath \else \i \fi{}o\ifmmode~\breve{g}\else
  \u{g}\fi{}lu}}, \bibinfo {author} {\bibfnamefont {L.~M.}\ \bibnamefont
  {Sandratskii}}, \bibinfo {author} {\bibfnamefont {P.}~\bibnamefont {Bruno}},\
  and\ \bibinfo {author} {\bibfnamefont {I.}~\bibnamefont {Galanakis}},\
  }\href@noop {} {\bibfield  {journal} {\bibinfo  {journal} {Phys. Rev. B}\
  }\textbf {\bibinfo {volume} {72}},\ \bibinfo {pages} {184415} (\bibinfo
  {year} {2005})}\BibitemShut {NoStop}%
\bibitem [{3-n()}]{3-note}%
  \BibitemOpen
  \href@noop {} {\bibinfo {title} {{In this paper the spin-orbit interaction is
  neglected and thus the Dzyaloshinskii-Moriya interaction or the magnetic
  anisotropies are not part of the discussion.}}}\BibitemShut {Stop}%
\bibitem [{\citenamefont {Iwashita}\ and\ \citenamefont
  {Ury\^u}(1976)}]{Iwashita:1976}%
  \BibitemOpen
  \bibfield  {author} {\bibinfo {author} {\bibfnamefont {T.}~\bibnamefont
  {Iwashita}}\ and\ \bibinfo {author} {\bibfnamefont {N.}~\bibnamefont
  {Ury\^u}},\ }\href@noop {} {\bibfield  {journal} {\bibinfo  {journal} {Phys.
  Rev. B}\ }\textbf {\bibinfo {volume} {14}},\ \bibinfo {pages} {3090}
  (\bibinfo {year} {1976})}\BibitemShut {NoStop}%
\bibitem [{\citenamefont {Ferriani}\ \emph {et~al.}(2008)\citenamefont
  {Ferriani}, \citenamefont {von Bergmann}, \citenamefont {Vedmedenko},
  \citenamefont {Heinze}, \citenamefont {Bode}, \citenamefont {Heide},
  \citenamefont {Bihlmayer}, \citenamefont {Bl\"ugel},\ and\ \citenamefont
  {Wiesendanger}}]{PhysRevLett.101.027201}%
  \BibitemOpen
  \bibfield  {author} {\bibinfo {author} {\bibfnamefont {P.}~\bibnamefont
  {Ferriani}}, \bibinfo {author} {\bibfnamefont {K.}~\bibnamefont {von
  Bergmann}}, \bibinfo {author} {\bibfnamefont {E.~Y.}\ \bibnamefont
  {Vedmedenko}}, \bibinfo {author} {\bibfnamefont {S.}~\bibnamefont {Heinze}},
  \bibinfo {author} {\bibfnamefont {M.}~\bibnamefont {Bode}}, \bibinfo {author}
  {\bibfnamefont {M.}~\bibnamefont {Heide}}, \bibinfo {author} {\bibfnamefont
  {G.}~\bibnamefont {Bihlmayer}}, \bibinfo {author} {\bibfnamefont
  {S.}~\bibnamefont {Bl\"ugel}},\ and\ \bibinfo {author} {\bibfnamefont
  {R.}~\bibnamefont {Wiesendanger}},\ }\href@noop {} {\bibfield  {journal}
  {\bibinfo  {journal} {Phys. Rev. Lett.}\ }\textbf {\bibinfo {volume} {101}},\
  \bibinfo {pages} {027201} (\bibinfo {year} {2008})}\BibitemShut {NoStop}%
\bibitem [{\citenamefont {Kurz}\ \emph {et~al.}(2001)\citenamefont {Kurz},
  \citenamefont {Bihlmayer}, \citenamefont {Hirai},\ and\ \citenamefont
  {Bl\"ugel}}]{Kurz3q}%
  \BibitemOpen
  \bibfield  {author} {\bibinfo {author} {\bibfnamefont {P.}~\bibnamefont
  {Kurz}}, \bibinfo {author} {\bibfnamefont {G.}~\bibnamefont {Bihlmayer}},
  \bibinfo {author} {\bibfnamefont {K.}~\bibnamefont {Hirai}},\ and\ \bibinfo
  {author} {\bibfnamefont {S.}~\bibnamefont {Bl\"ugel}},\ }\href@noop {}
  {\bibfield  {journal} {\bibinfo  {journal} {Phys. Rev. Lett.}\ }\textbf
  {\bibinfo {volume} {86}},\ \bibinfo {pages} {1106} (\bibinfo {year}
  {2001})}\BibitemShut {NoStop}%
\bibitem [{\citenamefont {Kittel}(1960)}]{Kittel:1960}%
  \BibitemOpen
  \bibfield  {author} {\bibinfo {author} {\bibfnamefont {C.}~\bibnamefont
  {Kittel}},\ }\href@noop {} {\bibfield  {journal} {\bibinfo  {journal} {Phys.
  Rev.}\ }\textbf {\bibinfo {volume} {120}},\ \bibinfo {pages} {335} (\bibinfo
  {year} {1960})}\BibitemShut {NoStop}%
\bibitem [{\citenamefont {Lines}(1972)}]{Lines:1972}%
  \BibitemOpen
  \bibfield  {author} {\bibinfo {author} {\bibfnamefont {M.}~\bibnamefont
  {Lines}},\ }\href@noop {} {\bibfield  {journal} {\bibinfo  {journal} {Solid
  State Commun.}\ }\textbf {\bibinfo {volume} {11}},\ \bibinfo {pages} {1615 }
  (\bibinfo {year} {1972})}\BibitemShut {NoStop}%
\bibitem [{\citenamefont {Bruno}(1993)}]{Bruno:1993}%
  \BibitemOpen
  \bibfield  {author} {\bibinfo {author} {\bibfnamefont {P.}~\bibnamefont
  {Bruno}},\ }\href@noop {} {\bibfield  {journal} {\bibinfo  {journal} {J.
  Magn. Magn. Mater.}\ }\textbf {\bibinfo {volume} {121}},\ \bibinfo {pages}
  {248 } (\bibinfo {year} {1993})},\ \bibinfo {note} {proceedings of the
  International Symposium on Magnetic Ultrathin Films, Multilayers and
  Surfaces}\BibitemShut {NoStop}%
\bibitem [{\citenamefont {Aharony}(1977)}]{0305-4470-10-3-011}%
  \BibitemOpen
  \bibfield  {author} {\bibinfo {author} {\bibfnamefont {A.}~\bibnamefont
  {Aharony}},\ }\href@noop {} {\bibfield  {journal} {\bibinfo  {journal} {J.
  Phys. A}\ }\textbf {\bibinfo {volume} {10}},\ \bibinfo {pages} {389}
  (\bibinfo {year} {1977})}\BibitemShut {NoStop}%
\bibitem [{\citenamefont {Sudano}\ and\ \citenamefont
  {Levy}(1978)}]{PhysRevB.18.5078}%
  \BibitemOpen
  \bibfield  {author} {\bibinfo {author} {\bibfnamefont {J.~J.}\ \bibnamefont
  {Sudano}}\ and\ \bibinfo {author} {\bibfnamefont {P.~M.}\ \bibnamefont
  {Levy}},\ }\href@noop {} {\bibfield  {journal} {\bibinfo  {journal} {Phys.
  Rev. B}\ }\textbf {\bibinfo {volume} {18}},\ \bibinfo {pages} {5078}
  (\bibinfo {year} {1978})}\BibitemShut {NoStop}%
\bibitem [{\citenamefont {Ury\^u}\ and\ \citenamefont
  {Friedberg}(1965)}]{PhysRev.140.A1803}%
  \BibitemOpen
  \bibfield  {author} {\bibinfo {author} {\bibfnamefont {N.}~\bibnamefont
  {Ury\^u}}\ and\ \bibinfo {author} {\bibfnamefont {S.~A.}\ \bibnamefont
  {Friedberg}},\ }\href@noop {} {\bibfield  {journal} {\bibinfo  {journal}
  {Phys. Rev.}\ }\textbf {\bibinfo {volume} {140}},\ \bibinfo {pages} {A1803}
  (\bibinfo {year} {1965})}\BibitemShut {NoStop}%
\bibitem [{\citenamefont {Rodbell}\ \emph {et~al.}(1963)\citenamefont
  {Rodbell}, \citenamefont {Jacobs}, \citenamefont {Owen},\ and\ \citenamefont
  {Harris}}]{Rodbell:1963}%
  \BibitemOpen
  \bibfield  {author} {\bibinfo {author} {\bibfnamefont {D.~S.}\ \bibnamefont
  {Rodbell}}, \bibinfo {author} {\bibfnamefont {I.~S.}\ \bibnamefont {Jacobs}},
  \bibinfo {author} {\bibfnamefont {J.}~\bibnamefont {Owen}},\ and\ \bibinfo
  {author} {\bibfnamefont {E.~A.}\ \bibnamefont {Harris}},\ }\href@noop {}
  {\bibfield  {journal} {\bibinfo  {journal} {Phys. Rev. Lett.}\ }\textbf
  {\bibinfo {volume} {11}},\ \bibinfo {pages} {10} (\bibinfo {year}
  {1963})}\BibitemShut {NoStop}%
\bibitem [{\citenamefont {Gaulin}\ and\ \citenamefont
  {Collins}(1986)}]{Gaulin:1986}%
  \BibitemOpen
  \bibfield  {author} {\bibinfo {author} {\bibfnamefont {B.~D.}\ \bibnamefont
  {Gaulin}}\ and\ \bibinfo {author} {\bibfnamefont {M.~F.}\ \bibnamefont
  {Collins}},\ }\href@noop {} {\bibfield  {journal} {\bibinfo  {journal} {Phys.
  Rev. B}\ }\textbf {\bibinfo {volume} {33}},\ \bibinfo {pages} {6287}
  (\bibinfo {year} {1986})}\BibitemShut {NoStop}%
\bibitem [{\citenamefont {Bhattacharjee}\ \emph {et~al.}(2006)\citenamefont
  {Bhattacharjee}, \citenamefont {Shenoy},\ and\ \citenamefont
  {Senthil}}]{Bhattacharjee:2006}%
  \BibitemOpen
  \bibfield  {author} {\bibinfo {author} {\bibfnamefont {S.}~\bibnamefont
  {Bhattacharjee}}, \bibinfo {author} {\bibfnamefont {V.~B.}\ \bibnamefont
  {Shenoy}},\ and\ \bibinfo {author} {\bibfnamefont {T.}~\bibnamefont
  {Senthil}},\ }\href@noop {} {\bibfield  {journal} {\bibinfo  {journal} {Phys.
  Rev. B}\ }\textbf {\bibinfo {volume} {74}},\ \bibinfo {pages} {092406}
  (\bibinfo {year} {2006})}\BibitemShut {NoStop}%
\bibitem [{\citenamefont {Fridman}\ and\ \citenamefont
  {Spirin}(2000)}]{Fridman:2000}%
  \BibitemOpen
  \bibfield  {author} {\bibinfo {author} {\bibfnamefont {Y.~A.}\ \bibnamefont
  {Fridman}}\ and\ \bibinfo {author} {\bibfnamefont {D.~V.}\ \bibnamefont
  {Spirin}},\ }\href@noop {} {\bibfield  {journal} {\bibinfo  {journal} {Low
  Temp. Phys.}\ }\textbf {\bibinfo {volume} {26}},\ \bibinfo {pages} {273}
  (\bibinfo {year} {2000})}\BibitemShut {NoStop}%
\bibitem [{\citenamefont {Lou}\ \emph {et~al.}(2000)\citenamefont {Lou},
  \citenamefont {Xiang},\ and\ \citenamefont {Su}}]{Lou:2000}%
  \BibitemOpen
  \bibfield  {author} {\bibinfo {author} {\bibfnamefont {J.}~\bibnamefont
  {Lou}}, \bibinfo {author} {\bibfnamefont {T.}~\bibnamefont {Xiang}},\ and\
  \bibinfo {author} {\bibfnamefont {Z.}~\bibnamefont {Su}},\ }\href@noop {}
  {\bibfield  {journal} {\bibinfo  {journal} {Phys. Rev. Lett.}\ }\textbf
  {\bibinfo {volume} {85}},\ \bibinfo {pages} {2380} (\bibinfo {year}
  {2000})}\BibitemShut {NoStop}%
\bibitem [{\citenamefont {Brown}(1975)}]{Brown:1975}%
  \BibitemOpen
  \bibfield  {author} {\bibinfo {author} {\bibfnamefont {H.~A.}\ \bibnamefont
  {Brown}},\ }\href@noop {} {\bibfield  {journal} {\bibinfo  {journal} {Phys.
  Rev. B}\ }\textbf {\bibinfo {volume} {11}},\ \bibinfo {pages} {4725}
  (\bibinfo {year} {1975})}\BibitemShut {NoStop}%
\bibitem [{\citenamefont {Takahashi}(1977)}]{takahashi}%
  \BibitemOpen
  \bibfield  {author} {\bibinfo {author} {\bibfnamefont {M.}~\bibnamefont
  {Takahashi}},\ }\href@noop {} {\bibfield  {journal} {\bibinfo  {journal} {J.
  Phys. C Solid State Phys.}\ }\textbf {\bibinfo {volume} {10}},\ \bibinfo
  {pages} {1289} (\bibinfo {year} {1977})}\BibitemShut {NoStop}%
\bibitem [{\citenamefont {MacDonald}\ \emph {et~al.}(1988)\citenamefont
  {MacDonald}, \citenamefont {Girvin},\ and\ \citenamefont
  {Yoshioka}}]{macdonald}%
  \BibitemOpen
  \bibfield  {author} {\bibinfo {author} {\bibfnamefont {A.~H.}\ \bibnamefont
  {MacDonald}}, \bibinfo {author} {\bibfnamefont {S.~M.}\ \bibnamefont
  {Girvin}},\ and\ \bibinfo {author} {\bibfnamefont {D.}~\bibnamefont
  {Yoshioka}},\ }\href@noop {} {\bibfield  {journal} {\bibinfo  {journal}
  {Phys. Rev. B}\ }\textbf {\bibinfo {volume} {37}},\ \bibinfo {pages} {9753}
  (\bibinfo {year} {1988})}\BibitemShut {NoStop}%
\bibitem [{\citenamefont {K{\"o}bler}\ \emph {et~al.}(1996)\citenamefont
  {K{\"o}bler}, \citenamefont {Mueller}, \citenamefont {Smardz}, \citenamefont
  {Maier}, \citenamefont {Fischer}, \citenamefont {Olefs},\ and\ \citenamefont
  {Zinn}}]{Kobler1996}%
  \BibitemOpen
  \bibfield  {author} {\bibinfo {author} {\bibfnamefont {U.}~\bibnamefont
  {K{\"o}bler}}, \bibinfo {author} {\bibfnamefont {R.}~\bibnamefont {Mueller}},
  \bibinfo {author} {\bibfnamefont {L.}~\bibnamefont {Smardz}}, \bibinfo
  {author} {\bibfnamefont {D.}~\bibnamefont {Maier}}, \bibinfo {author}
  {\bibfnamefont {K.}~\bibnamefont {Fischer}}, \bibinfo {author} {\bibfnamefont
  {B.}~\bibnamefont {Olefs}},\ and\ \bibinfo {author} {\bibfnamefont
  {W.}~\bibnamefont {Zinn}},\ }\href {https://doi.org/10.1007/s002570050153}
  {\bibfield  {journal} {\bibinfo  {journal} {Z. Phys. B}\ }\textbf {\bibinfo
  {volume} {100}},\ \bibinfo {pages} {497} (\bibinfo {year}
  {1996})}\BibitemShut {NoStop}%
\bibitem [{\citenamefont {Köbler}\ \emph {et~al.}(2001)\citenamefont
  {Köbler}, \citenamefont {Hoser}, \citenamefont {Englich}, \citenamefont
  {Snezhko}, \citenamefont {Kawakami}, \citenamefont {Beyss},\ and\
  \citenamefont {Fischer}}]{JPSJ.70.3089}%
  \BibitemOpen
  \bibfield  {author} {\bibinfo {author} {\bibfnamefont {U.}~\bibnamefont
  {Köbler}}, \bibinfo {author} {\bibfnamefont {A.}~\bibnamefont {Hoser}},
  \bibinfo {author} {\bibfnamefont {J.}~\bibnamefont {Englich}}, \bibinfo
  {author} {\bibfnamefont {A.}~\bibnamefont {Snezhko}}, \bibinfo {author}
  {\bibfnamefont {M.}~\bibnamefont {Kawakami}}, \bibinfo {author}
  {\bibfnamefont {M.}~\bibnamefont {Beyss}},\ and\ \bibinfo {author}
  {\bibfnamefont {K.}~\bibnamefont {Fischer}},\ }\href
  {https://doi.org/10.1143/JPSJ.70.3089} {\bibfield  {journal} {\bibinfo
  {journal} {J. Phys. Soc. Jpn.}\ }\textbf {\bibinfo {volume} {70}},\ \bibinfo
  {pages} {3089} (\bibinfo {year} {2001})}\BibitemShut {NoStop}%
\bibitem [{\citenamefont {Mryasov}\ \emph {et~al.}(1996)\citenamefont
  {Mryasov}, \citenamefont {Freeman},\ and\ \citenamefont
  {Liechtenstein}}]{doi:10.1063/1.361678}%
  \BibitemOpen
  \bibfield  {author} {\bibinfo {author} {\bibfnamefont {O.~N.}\ \bibnamefont
  {Mryasov}}, \bibinfo {author} {\bibfnamefont {A.~J.}\ \bibnamefont
  {Freeman}},\ and\ \bibinfo {author} {\bibfnamefont {A.~I.}\ \bibnamefont
  {Liechtenstein}},\ }\href@noop {} {\bibfield  {journal} {\bibinfo  {journal}
  {J. Appl. Phys.}\ }\textbf {\bibinfo {volume} {79}},\ \bibinfo {pages} {4805}
  (\bibinfo {year} {1996})}\BibitemShut {NoStop}%
\bibitem [{\citenamefont {Yoshida}\ \emph {et~al.}(2012)\citenamefont
  {Yoshida}, \citenamefont {Schr\"oder}, \citenamefont {Ferriani},
  \citenamefont {Serrate}, \citenamefont {Kubetzka}, \citenamefont {von
  Bergmann}, \citenamefont {Heinze},\ and\ \citenamefont
  {Wiesendanger}}]{silke}%
  \BibitemOpen
  \bibfield  {author} {\bibinfo {author} {\bibfnamefont {Y.}~\bibnamefont
  {Yoshida}}, \bibinfo {author} {\bibfnamefont {S.}~\bibnamefont {Schr\"oder}},
  \bibinfo {author} {\bibfnamefont {P.}~\bibnamefont {Ferriani}}, \bibinfo
  {author} {\bibfnamefont {D.}~\bibnamefont {Serrate}}, \bibinfo {author}
  {\bibfnamefont {A.}~\bibnamefont {Kubetzka}}, \bibinfo {author}
  {\bibfnamefont {K.}~\bibnamefont {von Bergmann}}, \bibinfo {author}
  {\bibfnamefont {S.}~\bibnamefont {Heinze}},\ and\ \bibinfo {author}
  {\bibfnamefont {R.}~\bibnamefont {Wiesendanger}},\ }\href@noop {} {\bibfield
  {journal} {\bibinfo  {journal} {Phys. Rev. Lett.}\ }\textbf {\bibinfo
  {volume} {108}},\ \bibinfo {pages} {087205} (\bibinfo {year}
  {2012})}\BibitemShut {NoStop}%
\bibitem [{\citenamefont {Heinze}\ \emph {et~al.}(2011)\citenamefont {Heinze},
  \citenamefont {von Bergmann}, \citenamefont {Menzel}, \citenamefont {Brede},
  \citenamefont {Kubetzka}, \citenamefont {Wiesendanger}, \citenamefont
  {Bihlmayer},\ and\ \citenamefont {Bl{\"u}gel}}]{heinze2011}%
  \BibitemOpen
  \bibfield  {author} {\bibinfo {author} {\bibfnamefont {S.}~\bibnamefont
  {Heinze}}, \bibinfo {author} {\bibfnamefont {K.}~\bibnamefont {von
  Bergmann}}, \bibinfo {author} {\bibfnamefont {M.}~\bibnamefont {Menzel}},
  \bibinfo {author} {\bibfnamefont {J.}~\bibnamefont {Brede}}, \bibinfo
  {author} {\bibfnamefont {A.}~\bibnamefont {Kubetzka}}, \bibinfo {author}
  {\bibfnamefont {R.}~\bibnamefont {Wiesendanger}}, \bibinfo {author}
  {\bibfnamefont {G.}~\bibnamefont {Bihlmayer}},\ and\ \bibinfo {author}
  {\bibfnamefont {S.}~\bibnamefont {Bl{\"u}gel}},\ }\href@noop {} {\bibfield
  {journal} {\bibinfo  {journal} {Nat. Phys.}\ }\textbf {\bibinfo {volume}
  {7}},\ \bibinfo {pages} {713} (\bibinfo {year} {2011})}\BibitemShut {NoStop}%
\bibitem [{\citenamefont {Al-Zubi}\ \emph {et~al.}(2011)\citenamefont
  {Al-Zubi}, \citenamefont {Bihlmayer},\ and\ \citenamefont
  {Blügel}}]{PSSB:PSSB201147090}%
  \BibitemOpen
  \bibfield  {author} {\bibinfo {author} {\bibfnamefont {A.}~\bibnamefont
  {Al-Zubi}}, \bibinfo {author} {\bibfnamefont {G.}~\bibnamefont {Bihlmayer}},\
  and\ \bibinfo {author} {\bibfnamefont {S.}~\bibnamefont {Blügel}},\
  }\href@noop {} {\bibfield  {journal} {\bibinfo  {journal} {Phys. Status
  Solidi (b)}\ }\textbf {\bibinfo {volume} {248}},\ \bibinfo {pages} {2242}
  (\bibinfo {year} {2011})}\BibitemShut {NoStop}%
\bibitem [{\citenamefont {Kr{\"o}nlein}\ \emph {et~al.}(2018)\citenamefont
  {Kr{\"o}nlein}, \citenamefont {Schmitt}, \citenamefont {Hoffmann},
  \citenamefont {Kemmer}, \citenamefont {Seubert}, \citenamefont {Vogt},
  \citenamefont {K{\"u}spert}, \citenamefont {B{\"o}hme}, \citenamefont
  {Alonazi}, \citenamefont {K{\"u}gel} \emph {et~al.}}]{kronlein}%
  \BibitemOpen
  \bibfield  {author} {\bibinfo {author} {\bibfnamefont {A.}~\bibnamefont
  {Kr{\"o}nlein}}, \bibinfo {author} {\bibfnamefont {M.}~\bibnamefont
  {Schmitt}}, \bibinfo {author} {\bibfnamefont {M.}~\bibnamefont {Hoffmann}},
  \bibinfo {author} {\bibfnamefont {J.}~\bibnamefont {Kemmer}}, \bibinfo
  {author} {\bibfnamefont {N.}~\bibnamefont {Seubert}}, \bibinfo {author}
  {\bibfnamefont {M.}~\bibnamefont {Vogt}}, \bibinfo {author} {\bibfnamefont
  {J.}~\bibnamefont {K{\"u}spert}}, \bibinfo {author} {\bibfnamefont
  {M.}~\bibnamefont {B{\"o}hme}}, \bibinfo {author} {\bibfnamefont
  {B.}~\bibnamefont {Alonazi}}, \bibinfo {author} {\bibfnamefont
  {J.}~\bibnamefont {K{\"u}gel}}, \emph {et~al.},\ }\href@noop {} {\bibfield
  {journal} {\bibinfo  {journal} {Phys. Rev. Lett.}\ }\textbf {\bibinfo
  {volume} {120}},\ \bibinfo {pages} {207202} (\bibinfo {year}
  {2018})}\BibitemShut {NoStop}%
\bibitem [{\citenamefont {Romming}\ \emph {et~al.}(2018)\citenamefont
  {Romming}, \citenamefont {Pralow}, \citenamefont {Kubetzka}, \citenamefont
  {Hoffmann}, \citenamefont {von Malottki}, \citenamefont {Meyer},
  \citenamefont {Dup\'e}, \citenamefont {Wiesendanger}, \citenamefont {von
  Bergmann},\ and\ \citenamefont {Heinze}}]{PhysRevLett.120.207201}%
  \BibitemOpen
  \bibfield  {author} {\bibinfo {author} {\bibfnamefont {N.}~\bibnamefont
  {Romming}}, \bibinfo {author} {\bibfnamefont {H.}~\bibnamefont {Pralow}},
  \bibinfo {author} {\bibfnamefont {A.}~\bibnamefont {Kubetzka}}, \bibinfo
  {author} {\bibfnamefont {M.}~\bibnamefont {Hoffmann}}, \bibinfo {author}
  {\bibfnamefont {S.}~\bibnamefont {von Malottki}}, \bibinfo {author}
  {\bibfnamefont {S.}~\bibnamefont {Meyer}}, \bibinfo {author} {\bibfnamefont
  {B.}~\bibnamefont {Dup\'e}}, \bibinfo {author} {\bibfnamefont
  {R.}~\bibnamefont {Wiesendanger}}, \bibinfo {author} {\bibfnamefont
  {K.}~\bibnamefont {von Bergmann}},\ and\ \bibinfo {author} {\bibfnamefont
  {S.}~\bibnamefont {Heinze}},\ }\href
  {https://doi.org/10.1103/PhysRevLett.120.207201} {\bibfield  {journal}
  {\bibinfo  {journal} {Phys. Rev. Lett.}\ }\textbf {\bibinfo {volume} {120}},\
  \bibinfo {pages} {207201} (\bibinfo {year} {2018})}\BibitemShut {NoStop}%
\bibitem [{4-n()}]{4-note}%
  \BibitemOpen
  \href@noop {} {\bibinfo {title} {{We would like to note that the nomenclature
  three-spin interactions has recently drawn criticism as the interaction
  indicates an interaction of three spins, which would lack time-inversion
  symmetry. In the context of this paper we use it synonymously to
  four-spin-three-site interaction.}}}\BibitemShut {Stop}%
\bibitem [{\citenamefont {Iwashita}\ and\ \citenamefont
  {Uryû}(1974)}]{doi:10.1143/JPSJ.36.48}%
  \BibitemOpen
  \bibfield  {author} {\bibinfo {author} {\bibfnamefont {T.}~\bibnamefont
  {Iwashita}}\ and\ \bibinfo {author} {\bibfnamefont {N.}~\bibnamefont
  {Uryû}},\ }\href@noop {} {\bibfield  {journal} {\bibinfo  {journal} {J.
  Phys. Soc. Jpn.}\ }\textbf {\bibinfo {volume} {36}},\ \bibinfo {pages} {48}
  (\bibinfo {year} {1974})}\BibitemShut {NoStop}%
\bibitem [{\citenamefont {Iwashita}\ and\ \citenamefont
  {Uryû}(1979)}]{doi:10.1143/JPSJ.47.786}%
  \BibitemOpen
  \bibfield  {author} {\bibinfo {author} {\bibfnamefont {T.}~\bibnamefont
  {Iwashita}}\ and\ \bibinfo {author} {\bibfnamefont {N.}~\bibnamefont
  {Uryû}},\ }\href@noop {} {\bibfield  {journal} {\bibinfo  {journal} {J.
  Phys. Soc. Jpn.}\ }\textbf {\bibinfo {volume} {47}},\ \bibinfo {pages} {786}
  (\bibinfo {year} {1979})}\BibitemShut {NoStop}%
\bibitem [{\citenamefont {Iwashita}\ and\ \citenamefont
  {Uryǔ}(1980)}]{IWASHITA198064}%
  \BibitemOpen
  \bibfield  {author} {\bibinfo {author} {\bibfnamefont {T.}~\bibnamefont
  {Iwashita}}\ and\ \bibinfo {author} {\bibfnamefont {N.}~\bibnamefont
  {Uryǔ}},\ }\href@noop {} {\bibfield  {journal} {\bibinfo  {journal} {Phys.
  Lett. A}\ }\textbf {\bibinfo {volume} {76}},\ \bibinfo {pages} {64 }
  (\bibinfo {year} {1980})}\BibitemShut {NoStop}%
\bibitem [{\citenamefont {Kobayashi}\ \emph {et~al.}(1968)\citenamefont
  {Kobayashi}, \citenamefont {Tsujikawa},\ and\ \citenamefont
  {Kimura}}]{19681169}%
  \BibitemOpen
  \bibfield  {author} {\bibinfo {author} {\bibfnamefont {H.}~\bibnamefont
  {Kobayashi}}, \bibinfo {author} {\bibfnamefont {I.}~\bibnamefont
  {Tsujikawa}},\ and\ \bibinfo {author} {\bibfnamefont {I.}~\bibnamefont
  {Kimura}},\ }\href {https://doi.org/10.1143/JPSJ.24.1169} {\bibfield
  {journal} {\bibinfo  {journal} {J. Phys. Soc. Jpn.}\ }\textbf {\bibinfo
  {volume} {24}},\ \bibinfo {pages} {1169} (\bibinfo {year}
  {1968})}\BibitemShut {NoStop}%
\bibitem [{\citenamefont {Bastardis}\ \emph {et~al.}(2007)\citenamefont
  {Bastardis}, \citenamefont {Guih\'ery},\ and\ \citenamefont
  {de~Graaf}}]{Bastardis:2007}%
  \BibitemOpen
  \bibfield  {author} {\bibinfo {author} {\bibfnamefont {R.}~\bibnamefont
  {Bastardis}}, \bibinfo {author} {\bibfnamefont {N.}~\bibnamefont
  {Guih\'ery}},\ and\ \bibinfo {author} {\bibfnamefont {C.}~\bibnamefont
  {de~Graaf}},\ }\href@noop {} {\bibfield  {journal} {\bibinfo  {journal}
  {Phys. Rev. B}\ }\textbf {\bibinfo {volume} {76}},\ \bibinfo {pages} {132412}
  (\bibinfo {year} {2007})}\BibitemShut {NoStop}%
\bibitem [{\citenamefont {M{\"u}ller-Hartmann}\ \emph
  {et~al.}(1997)\citenamefont {M{\"u}ller-Hartmann}, \citenamefont
  {K{\"o}bler},\ and\ \citenamefont {Smardz}}]{muller1997}%
  \BibitemOpen
  \bibfield  {author} {\bibinfo {author} {\bibfnamefont {E.}~\bibnamefont
  {M{\"u}ller-Hartmann}}, \bibinfo {author} {\bibfnamefont {U.}~\bibnamefont
  {K{\"o}bler}},\ and\ \bibinfo {author} {\bibfnamefont {L.}~\bibnamefont
  {Smardz}},\ }\href@noop {} {\bibfield  {journal} {\bibinfo  {journal} {J.
  Magn. Magn. Mater.}\ }\textbf {\bibinfo {volume} {173}},\ \bibinfo {pages}
  {133} (\bibinfo {year} {1997})}\BibitemShut {NoStop}%
\bibitem [{\citenamefont {L{\"o}wdin}(1951)}]{lowdin}%
  \BibitemOpen
  \bibfield  {author} {\bibinfo {author} {\bibfnamefont {P.-O.}\ \bibnamefont
  {L{\"o}wdin}},\ }\href@noop {} {\bibfield  {journal} {\bibinfo  {journal} {J.
  Chem. Phys.}\ }\textbf {\bibinfo {volume} {19}},\ \bibinfo {pages} {1396}
  (\bibinfo {year} {1951})}\BibitemShut {NoStop}%
\bibitem [{\citenamefont {Winkler}(2003)}]{winkler}%
  \BibitemOpen
  \bibfield  {author} {\bibinfo {author} {\bibfnamefont {R.}~\bibnamefont
  {Winkler}},\ }\href@noop {} {\emph {\bibinfo {title} {Spin-orbit coupling
  effects in two-dimensional electron and hole systems}}},\ Vol.\ \bibinfo
  {volume} {191}\ (\bibinfo  {publisher} {Springer},\ \bibinfo {year}
  {2003})\BibitemShut {NoStop}%
\bibitem [{\citenamefont {Schrieffer}\ and\ \citenamefont
  {Wolff}(1966)}]{Schrieffer:66}%
  \BibitemOpen
  \bibfield  {author} {\bibinfo {author} {\bibfnamefont {J.~R.}\ \bibnamefont
  {Schrieffer}}\ and\ \bibinfo {author} {\bibfnamefont {P.~A.}\ \bibnamefont
  {Wolff}},\ }\href@noop {} {\bibfield  {journal} {\bibinfo  {journal} {Phys.
  Rev.}\ }\textbf {\bibinfo {volume} {149}},\ \bibinfo {pages} {491} (\bibinfo
  {year} {1966})}\BibitemShut {NoStop}%
\bibitem [{\citenamefont {Bravyi}\ \emph {et~al.}(2011)\citenamefont {Bravyi},
  \citenamefont {DiVincenzo},\ and\ \citenamefont {Loss}}]{Bravyi2011}%
  \BibitemOpen
  \bibfield  {author} {\bibinfo {author} {\bibfnamefont {S.}~\bibnamefont
  {Bravyi}}, \bibinfo {author} {\bibfnamefont {D.~P.}\ \bibnamefont
  {DiVincenzo}},\ and\ \bibinfo {author} {\bibfnamefont {D.}~\bibnamefont
  {Loss}},\ }\href {https://doi.org/10.1016/j.aop.2011.06.004} {\bibfield
  {journal} {\bibinfo  {journal} {Ann. Phys. (N. Y.)}\ }\textbf {\bibinfo
  {volume} {326}},\ \bibinfo {pages} {2793} (\bibinfo {year}
  {2011})}\BibitemShut {NoStop}%
\bibitem [{\citenamefont {Hubbard}(1963)}]{hubbard}%
  \BibitemOpen
  \bibfield  {author} {\bibinfo {author} {\bibfnamefont {J.}~\bibnamefont
  {Hubbard}},\ }\href@noop {} {\bibfield  {journal} {\bibinfo  {journal} {Proc.
  Royal Soc. A}\ }\textbf {\bibinfo {volume} {276}},\ \bibinfo {pages} {238}
  (\bibinfo {year} {1963})}\BibitemShut {NoStop}%
\bibitem [{\citenamefont {Hubbard}(1964{\natexlab{a}})}]{hubbard2}%
  \BibitemOpen
  \bibfield  {author} {\bibinfo {author} {\bibfnamefont {J.}~\bibnamefont
  {Hubbard}},\ }\href@noop {} {\bibfield  {journal} {\bibinfo  {journal} {Proc.
  Royal Soc. A}\ }\textbf {\bibinfo {volume} {277}},\ \bibinfo {pages} {237}
  (\bibinfo {year} {1964}{\natexlab{a}})}\BibitemShut {NoStop}%
\bibitem [{\citenamefont {Hubbard}(1964{\natexlab{b}})}]{hubbard3}%
  \BibitemOpen
  \bibfield  {author} {\bibinfo {author} {\bibfnamefont {J.}~\bibnamefont
  {Hubbard}},\ }\href@noop {} {\bibfield  {journal} {\bibinfo  {journal} {Proc.
  Royal Soc. A}\ }\textbf {\bibinfo {volume} {281}},\ \bibinfo {pages} {401}
  (\bibinfo {year} {1964}{\natexlab{b}})}\BibitemShut {NoStop}%
\bibitem [{1-n()}]{1-note-t2}%
  \BibitemOpen
  \href@noop {} {\bibinfo {title} {{We use $t_{i,\alpha,j,\alpha'}'$ instead of
  only $t_{i,\alpha,j,\alpha'}$ to emphasize the two different types of
  hoppings. This notation will also be used later on in the
  paper.}}}\BibitemShut {Stop}%
\bibitem [{2-n()}]{2-note-ferro}%
  \BibitemOpen
  \href@noop {} {\bibinfo {title} {{This ensures that a ferromagnetic alignment
  (either all spins up or all spins down) gains no energy in higher-order. This
  agrees well with the underlying Hubbard model where none of the hopping terms
  have any effect on those ferromagnetic states as no hopping is
  possible.}}}\BibitemShut {Stop}%
\bibitem [{\citenamefont {Hardrat}\ \emph {et~al.}(2009)\citenamefont
  {Hardrat}, \citenamefont {Al-Zubi}, \citenamefont {Ferriani}, \citenamefont
  {Bl\"ugel}, \citenamefont {Bihlmayer},\ and\ \citenamefont
  {Heinze}}]{PhysRevB.79.094411}%
  \BibitemOpen
  \bibfield  {author} {\bibinfo {author} {\bibfnamefont {B.}~\bibnamefont
  {Hardrat}}, \bibinfo {author} {\bibfnamefont {A.}~\bibnamefont {Al-Zubi}},
  \bibinfo {author} {\bibfnamefont {P.}~\bibnamefont {Ferriani}}, \bibinfo
  {author} {\bibfnamefont {S.}~\bibnamefont {Bl\"ugel}}, \bibinfo {author}
  {\bibfnamefont {G.}~\bibnamefont {Bihlmayer}},\ and\ \bibinfo {author}
  {\bibfnamefont {S.}~\bibnamefont {Heinze}},\ }\href@noop {} {\bibfield
  {journal} {\bibinfo  {journal} {Phys. Rev. B}\ }\textbf {\bibinfo {volume}
  {79}},\ \bibinfo {pages} {094411} (\bibinfo {year} {2009})}\BibitemShut
  {NoStop}%
\end{thebibliography}
\end{document}